\pdfoutput=1
\documentclass[acmsmall,screen,nonacm]{acmart}
\AtBeginDocument{%
  }

\usepackage{wrapfig}
\usepackage{xspace}
\usepackage{algorithm}
\usepackage{algpseudocode}
\usepackage{multirow}
\newcommand{\ourmodel}{\textsc{PAUSE}\xspace}
\newcommand{\gain}[1]{\textcolor{green!50!black}{#1}}
\newcommand{\tool}{\textsc{PAUSE}\xspace}

\usepackage[final]{labreview}

\newcommand{\commentout}[1]{} 
\definecolor{pausegray}{gray}{0.93}
\usepackage{xcolor}
\newcommand{\rqanswer}[1]{%
  \par\smallskip\noindent
  \fcolorbox{black!25}{pausegray}{%
    \parbox{\dimexpr\linewidth-2\fboxsep-2\fboxrule}{%
      \small\setlength{\parfillskip}{0pt plus 0.5\linewidth}%
      \textbf{Answer.}\ #1}}%
  \par\smallskip}

\begin{document}
\raggedbottom
\setlength{\textfloatsep}{10pt plus 2pt minus 2pt}
\setlength{\floatsep}{8pt plus 2pt minus 2pt}
\setlength{\intextsep}{8pt plus 2pt minus 2pt}
\setlength{\parfillskip}{0pt plus 0.5\linewidth}
\finalhyphendemerits=1000000

\title[Search-Free Steering of Coding Agents]{PAUSE: Search-Free Steering of Coding Agents Before Their First Source-Code Edit}
\author{Viral Patel}
\affiliation{%
  \institution{Mohamed bin Zayed University of Artificial Intelligence}
  \city{Abu Dhabi}
  \country{United Arab Emirates}}

\author{Boyin Tan}
\affiliation{%
  \institution{Mohamed bin Zayed University of Artificial Intelligence}
  \city{Abu Dhabi}
  \country{United Arab Emirates}}

\author{Jason Xue}
\affiliation{%
  \institution{Mohamed bin Zayed University of Artificial Intelligence}
  \city{Abu Dhabi}
  \country{United Arab Emirates}}

\author{Youcheng Sun}
\affiliation{%
  \institution{Mohamed bin Zayed University of Artificial Intelligence}
  \city{Abu Dhabi}
  \country{United Arab Emirates}}

\renewcommand{\shortauthors}{Patel et al.}

\begin{abstract}
Coding agents can invest substantial effort in repository exploration and testing yet still produce unsuccessful patches. Existing online steering methods commonly review execution trajectories periodically or invoke an advisor when a monitor detects behavioral problems. Frequent reviews are costly, and the absence of anomalous behavior does not mean that an agent is on the right track. To bridge this gap, we propose \tool{}, a framework for steering coding agents with a single planning intervention.
We find that agents reach their first source-code edit at a median of 60\% of trajectory progress, with the preceding work largely devoted to repository exploration, bug reproduction, and testing. This motivates intercepting the request before execution, so that the agent's accumulated investigation can guide the repair it is about to implement.
Because the first source-code edit directly identifies when to intervene, \tool{} locates this point without a learned monitor, a search over candidate checkpoints, or an extra model call---a property we call \emph{search-free}. At this point, a separate \emph{Planner} reads the task and pre-edit transcript to produce one repair plan. The same agent, the \emph{Executor}, resumes with this plan, retaining its accumulated context and workspace. \tool{} applies both online, during a live run, and offline, from a recorded trajectory prefix.

We evaluate \tool{} on SWE-bench Verified and SWE-bench Multilingual with two Executors and three agent harnesses. With Qwen3-coder as Executor and GPT-5.6 Sol as Planner, online steering in OpenHands improves resolution on Verified by 16.8\%, and online steering in SWE-agent improves resolution on Multilingual by 14.7\%. In SWE-agent, against LivePlan, which invokes an advisor through rule-based monitoring, \tool{} is more effective at a similar Planner cost, improving resolution by 3.2\% and 3.7\% on Verified and Multilingual, respectively. Against SICA, which reviews the trajectory periodically, \tool{} remains comparable on both benchmarks while incurring about 11\% and 15\% of SICA's Planner cost on Verified and Multilingual, respectively. Offline, \tool{} recovers 35.4\% of the eligible failed Verified tasks, and resuming from the recorded prefix uses 33--41\% fewer Executor tokens than restarting. These results show that a single planning intervention at the first source-code edit is more effective than rule-based monitoring and competitive with periodic oversight at a small fraction of its cost.
\end{abstract}



\maketitle

\section{Introduction}
\label{sec:introduction}

LLM-based coding agents tackle repository-level issues by inspecting code, making edits, and running commands and tests through an interactive tool interface~\cite{yang2024sweagent,wang2024openhands,Bouzenia2024RepairAgentAA}.
Existing repair systems gather relevant context through localization and structural code search~\cite{xia2024agentless,zhang2024autocoderover}, and infer code intent to guide repair~\cite{Ruan2024SpecRoverCI}.
However, agents can still struggle to turn this investigation into a correct repair~\cite{yang2024sweagent}.
This motivates \emph{online steering}, which uses the context accumulated during execution to guide the agent's subsequent repair decisions~\cite{gandhi2025swe,Nanda2026WinkRF,Liu2025ProcessCentricAO}.

The timing of guidance shapes both the effectiveness and cost of online steering~\cite{liu2026liveplan}.
As shown in Table~\ref{tab:steering-principles}, existing methods decide when to guide either through repeated model-based reviews or by waiting for a detected problem.
With \emph{periodic model-based review}, methods such as SWE-PRM~\cite{gandhi2025swe} and SICA~\cite{robeyns2025sica} repeatedly assess ongoing execution at regular step or time intervals to determine whether guidance is needed.
Each review incurs additional inference cost, while longer intervals can delay useful feedback until the agent has taken further steps along an ineffective path~\cite{liu2026liveplan}.
A fixed schedule also need not align with the point at which accumulated investigation is ready to inform a code change.
To reduce review overhead, LivePlan uses \emph{rule-based checks} to invoke an advisor when signals such as repetition, stagnation, or skipped phases are detected~\cite{liu2026liveplan}.
This reduces monitoring cost, but intervention remains limited by the coverage of predefined signals, which LivePlan acknowledges may be incomplete~\cite{liu2026liveplan}.
Thus, the absence of a behavioral alarm does not establish that the repair strategy is correct: an agent may follow an orderly workflow while preparing an ineffective change.

\begin{table}[t]
\centering
\small
\caption{Online steering methods compared by how they decide when to guide.
\emph{Trigger}: the event that starts an intervention.
\emph{Timing model call}: whether a model must be called to decide when to intervene.
\emph{Needs detected problem}: whether guidance is given only after a problem in the trajectory is detected.
\emph{Interventions}: how many times guidance is inserted into a single run.}
\label{tab:steering-principles}
\setlength{\tabcolsep}{2.5pt}
\begin{tabular}{@{}llccc@{}}
\toprule
\textbf{Method} & \textbf{Trigger} & \textbf{Timing model call} & \textbf{Needs detected problem} & \textbf{Interventions} \\
\midrule
SWE-PRM~\cite{gandhi2025swe} & Fixed step interval & Yes & Yes & Repeated \\
SICA~\cite{robeyns2025sica} & Overseer-chosen interval & Yes & Yes & Repeated \\
LivePlan~\cite{liu2026liveplan} & Rule-detected anomaly & No & Yes & Repeated \\
\textbf{\ourmodel{} (Ours)} & \textbf{First source-code edit} & \textbf{No} & \textbf{No} & \textbf{One} \\
\bottomrule
\end{tabular}
\end{table}

We therefore seek a moment to guide repair using the agent's accumulated investigation, without repeated trajectory reviews or waiting for a behavioral alarm.

\begin{wrapfigure}{r}{0.52\linewidth}
  \vspace{-0.7\baselineskip}
  \centering
  \includegraphics[width=\linewidth]{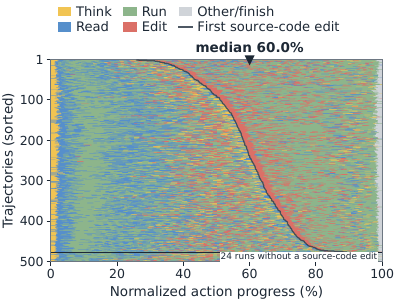}
  \caption{Actions in 500 Qwen3-coder runs on SWE-bench Verified. The curve marks first source-code edits (median 60.0\%); 24 runs without one lie below the divider.}
  \label{fig:checkpoint-position-distribution}
  \Description{A heatmap of 500 trajectories over normalized action progress from 0 to 100 percent. Yellow denotes think actions, blue file reads, green shell commands, red file-editor actions, and gray other or finish actions. The first 476 rows are sorted by their first source-code edit, traced by a black curve; a marker on the top edge labels their median position, 60.0 percent. The final 24 rows have no source-code edit and appear below a horizontal divider.}
  \vspace{-0.6\baselineskip}
\end{wrapfigure}

\paragraph{Why the first source-code edit?} To identify such a moment, we analyze 500 recorded Qwen3-coder trajectories on SWE-bench Verified.
We define the \emph{first source-code edit} as the first file-editor request targeting a source-code file of the repository.
We classify recorded actions into five categories: explicit reasoning, file inspection, shell command execution, file editing, and other operations, including termination.
Figure~\ref{fig:checkpoint-position-distribution} visualizes these sequences over normalized action progress and orders trajectories by their first source-code edit; edits to scratch files or reproduction scripts outside the designated source-code paths are excluded.
Among the 476 trajectories containing such a request, its median position is 60.0\% of action progress, and 80.5\% occur after the midpoint.
Before this point, command execution is the most common action type and file reads account for 31\% of actions, reflecting substantial repository investigation before the proposed source change.
This observation motivates intervening before the proposed source-code edit executes, when accumulated evidence is available to inform the repair and the proposed change remains open to revision.
Our experiments confirm that this point balances the two: an earlier intervention gives the Planner too little of the investigation to ground a plan, whereas a later one arrives after the agent has committed to and built on its own changes, which the plan must then overturn.

Prior work uses explicit planning to coordinate repository edits~\cite{Bairi2023CodePlanRC,Tao2024MAGISLM} and explores planning and execution roles in model collaboration~\cite{lee2026cope,gandhi2025empirical}.
To guide repair at this point, we introduce \textbf{\ourmodel{}}, a framework that uses the first source-code edit to trigger a single planning intervention.
The request identifies the intervention point directly from the agent's proposed action and its target path, without a learned monitor, a search over candidate checkpoints, or an additional model call to locate the point; we call this property \emph{search-free}.
A separate \emph{Planner} reads the task and the coding agent's preceding transcript, including inspected code and execution feedback, and returns one natural-language repair plan.
The same coding agent, the \emph{Executor}, receives this plan and continues from its existing conversation and workspace, retaining its investigation and responsibility for editing and testing without updating its underlying language model.

\ourmodel{} supports online steering and offline recovery.
Online, it blocks the first source-code edit during a live run and adds the Planner's guidance before resuming the Executor from its live state.
Offline, it restores an existing failed trajectory to a pre-edit state, adds a plan based on the recorded prefix, and resumes the same Executor with its remaining budget.

We evaluate \ourmodel{} on SWE-bench Verified~\cite{jimenez2024swebench} and SWE-bench Multilingual~\cite{yang2025swesmith}, which spans nine programming languages, using two Executors and three agent harnesses.
With Qwen3-coder as Executor and GPT-5.6 Sol as Planner in OpenHands, online steering improves resolution on Verified by 16.8\%.
In SWE-agent, it improves resolution by 14.8\% and 14.7\% on Verified and Multilingual, respectively.
Against LivePlan, which invokes an advisor through rule-based monitoring, \ourmodel{} is more effective at a similar Planner cost, improving resolution by 3.2\% and 3.7\% on Verified and Multilingual, respectively.
Against SICA, which reviews the trajectory periodically, \ourmodel{} remains comparable on both benchmarks while incurring about 11\% and 15\% of SICA's Planner cost on Verified and Multilingual, respectively.
Offline, \ourmodel{} recovers 79 of 223 eligible failed Verified tasks, a recovery rate of 35.4\%.
These results show that a single planning intervention at the first source-code edit is more effective than rule-based monitoring and competitive with periodic oversight at a small fraction of its cost.

Our contributions are:
\begin{itemize}
  \item An empirical characterization of the first source-code edit as an opportunity to guide repair~after substantial repository investigation.
  \item \ourmodel{}, a search-free framework that gives the same Executor one plan while preserving its context and workspace, supporting online steering and offline recovery.
  \item An evaluation across two benchmarks, two Executors, and three harnesses, showing that \ourmodel{} is more effective than rule-based monitoring at a similar Planner cost and is competitive with periodic oversight at 11--15\% of its Planner cost, with controlled studies of intervention timing, planning guidance, and state reuse.
\end{itemize}

\section{Related Work}
\label{sec:related-work}

\subsection{Repository-Level Repair and Planning}

SWE-bench frames repository-level repair as resolving real GitHub issues through changes to an existing codebase~\cite{jimenez2024swebench}.
SWE-agent equips language models with interfaces for navigating repositories, editing files, and using execution feedback~\cite{yang2024sweagent}.
OpenHands provides an extensible platform in which agents perform software development through code execution and other tools~\cite{wang2024openhands}.
CodeAgent integrates retrieval, navigation, implementation, and testing tools for repository-level coding~\cite{Zhang2024CodeAgentEC}, while RepoGraph and LingmaAgent use repository graphs to support context gathering and exploration~\cite{Ouyang2024RepoGraphEA,Ma2024AlibabaLI}.
Alongside these interactive agents, Agentless separates localization, repair, and patch validation into a structured pipeline~\cite{xia2024agentless}.
SHERLOC further studies dedicated diagnostic exploration, transferring localization findings to a downstream repair agent~\cite{tamoyan2026sherloc}.
Training-based approaches include SWE-Fixer's specialized retrieval and editing modules~\cite{Xie2025SWEFixerTO} and SWE-Gym's agent and verifier training in executable repository environments~\cite{Pan2024TrainingSE}.

Explicit planning also structures reasoning and collaboration.
Plan-and-Solve decomposes problems before solving them~\cite{Wang2023PlanandSolvePI}, and Plan-and-Act separates high-level planning from environment actions~\cite{Erdogan2025PlanandActIP}.
MetaGPT organizes software development through role-specific procedures and intermediate artifacts~\cite{Hong2023MetaGPTMP}, while Self-Collaboration coordinates analyst, coder, and tester roles~\cite{Dong2023SelfCollaborationCG}.
Within repository repair, planning can compare candidate solutions before implementation.
SWE-Manager trains an 8B model to select among competing repair proposals and synthesize a golden proposal for implementation~\cite{tan2026swemanager}.
Its P2A framework separates proposal generation, managerial selection and synthesis, and implementation into distinct roles~\cite{tan2026swemanager}.
\ourmodel{} instead obtains one plan from the investigation already accumulated by an executing agent, using its first source-code edit to determine when guidance is provided.
The same Executor then continues with its retained conversation and workspace.

\subsection{Online Steering}

Online steering introduces guidance while an agent is still carrying out its task.
SWE-PRM periodically reviews a recent trajectory window and produces natural-language feedback grounded in a taxonomy of execution inefficiencies~\cite{gandhi2025swe}.
SICA includes an asynchronous overseer that examines execution state at regular intervals or key events and can notify or stop an agent when it detects problematic behavior~\cite{robeyns2025sica}.
LivePlan separates detection from advice: a deterministic monitor checks for signals such as repetition, stagnation, and skipped phases, then requests corrective guidance when a signal fires~\cite{liu2026liveplan}.
These designs make the intervention decision depend on periodic assessment or detected behavioral conditions.
\ourmodel{} uses a different trigger: the first source-code edit identifies an opportunity to guide the repair regardless of whether the preceding trajectory exhibits a behavioral alarm.
This enables a single planning intervention without repeated model-based reviews or a search for an intervention point.

\subsection{Trajectory Reuse and Reflection}

Search-based approaches explore alternative reasoning and action sequences.
Tree of Thoughts searches over alternative reasoning paths~\cite{Yao2023TreeOT}, while Language Agent Tree Search combines tree search with environment feedback and reflection~\cite{Zhou2023LanguageAT}.
For code generation, CodeTree searches over coding strategies, candidate programs, and revisions using execution and agent feedback~\cite{Li2024CodeTreeAT}.

Within repository repair, reusing intermediate work can reduce exploration repeated across attempts.
SWE-Search combines Monte Carlo tree search with iterative refinement to explore alternative repair trajectories~\cite{antoniades2025swesearch}.
SWE-Replay retains an archive of sampled trajectories and selects intermediate steps from which to restore state and generate new continuations~\cite{ding2026swereplay}.
DARS resamples actions at branching points using prior execution feedback~\cite{Aggarwal2025DARSDA}, while SE-Agent revises, recombines, and refines existing trajectories~\cite{Lin2025SEAgentST}.
These methods use intermediate states to diversify the search for a successful repair.
Offline \ourmodel{} reuses a recorded prefix to support a different operation: inserting one plan at a checkpoint determined by the first source-code edit and continuing with the remaining execution budget.

Another approach uses verbal feedback to guide later attempts.
Reflexion stores verbal feedback from prior attempts in an episodic memory to guide subsequent trials~\cite{shinn2023reflexion}.
Self-Refine iterates between feedback generation and output revision using the same language model~\cite{madaan2023selfrefine}.
Tool-grounded feedback supports related correction mechanisms: CRITIC verifies and revises outputs through tool interaction~\cite{gou2024critic}, while Self-Debugging and LDB use program execution evidence to diagnose and revise generated code~\cite{chen2023teaching,Zhong2024DebugLA}.
ChatRepair retains candidate patches and test feedback within a repair conversation to inform subsequent attempts~\cite{Xia2023AutomatedPR}.

Execution feedback also supports candidate selection: CodeT uses generated tests and execution agreement~\cite{Chen2022CodeTCG}, whereas LEVER learns to score candidate programs from their inputs, code, and execution results~\cite{Ni2023LEVERLT}.
\ourmodel{} places the planning call within an existing coding trajectory and retains the Executor's conversation and workspace when delivering guidance.
The Planner reasons over the recorded evidence in a separate invocation, which may use the Executor's model.

\section{Method}
\label{sec:method}

\subsection{Problem Formulation}
\label{sec:formulation}

A repository-level repair task consists of an issue description $d$ and an
initial repository snapshot $R_0$~\cite{jimenez2024swebench}. A coding agent, which we call the
\emph{Executor}, investigates the issue, modifies files, and tests its changes
through an execution harness~\cite{yang2024sweagent,wang2024openhands}.
Reasoning-and-acting agents interleave model decisions with observations
from the environment~\cite{yao2023react}. After $t$ completed actions, the Executor's conversation
$h_t$ contains the system prompt, the task, and the recorded actions and
observations; $W_t$ is the resulting workspace. From this state, the
Executor proposes its next action $a_{t+1}$ for the harness to execute.

Online steering aims to improve the final repair during execution. It must
determine when to intervene and what guidance to provide using the task and
available evidence. We call the model supplying repair guidance the \emph{Planner}.
The available evidence includes inspected code, command outputs, and the
Executor's own test results. Benchmark evaluator feedback, hidden tests,
reference patches, and future actions are excluded from the planning input.

\subsection{Pre-Edit Intervention}
\label{sec:boundary}

The investigation preceding a source-code edit provides evidence for
guidance while the proposed change remains open to revision.
\ourmodel{} uses the first source-code edit to trigger
a single planning intervention, without repeated
trajectory assessment or a search over intervention points.

\begin{figure}[!htbp]
\centering
\includegraphics[width=\linewidth]{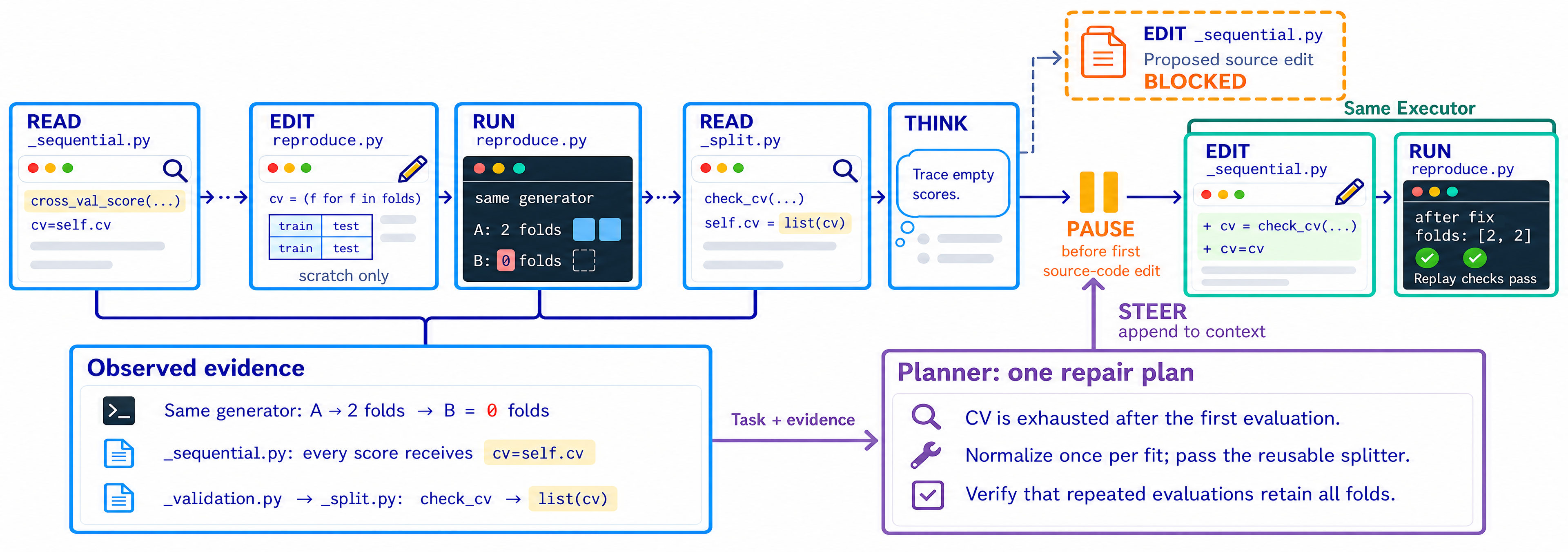}
\caption{Overview of \ourmodel{}. The Executor investigates the repository
by reading code, writing scratch scripts, and running them. When the Executor
issues its first source-code edit, \ourmodel{} blocks the request and passes the
task and the observed evidence to the Planner. The Planner returns one repair
plan, which \ourmodel{} appends to the Executor's conversation, and the same
Executor resumes with its context and workspace preserved.}
\Description{A reconstructed coding trajectory with small tool windows, bold action borders, and ellipses between representative investigation steps. Source reads and a scratch reproduction show that successive evaluations using the same cross-validation generator obtain two folds and then zero folds. Candidate scoring in the sequential feature selector repeatedly passes the original cv object through validation and splitter normalization. PAUSE blocks the first source-code edit. A lavender Planner panel combines these observations into a plan to normalize CV once per fit and propagate the reusable splitter to each candidate. A purple arrow appends the plan to context; the same Executor edits the sequential feature selector and checks that both evaluations retain two folds. The displayed fold counts come from local execution of the actual splitter functions.}
\label{fig:pause-overview}
\end{figure}

As illustrated in Figure~\ref{fig:pause-overview}, \ourmodel{} blocks the
request and asks the Planner to produce a repair plan from the preceding
evidence. It appends this plan to the same Executor's conversation, allowing
the agent to continue with its context and workspace preserved.

We identify a source-code edit by a file-editor request whose target is
a source-code file of the repository. Let $\Pi$ be the set of source-code
paths in the repository, fixed before execution. Writing
$\pi(a)$ for the target path, we define the predicate and the first source-code edit as
\begin{align}
q(a) &= \bigl[\,a\text{ is an editor request}\,\bigr]
        \wedge \bigl[\,\pi(a)\in\Pi\,\bigr],
\label{eq:source-edit}\\
k^\ast &= \min\{\,t:q(a_t)\,\}.
\label{eq:boundary}
\end{align}
The predicate can be evaluated before the requested edit executes, using
only the action type and target path. No model call or future observation
is needed to locate the boundary; this is its \emph{search-free} property.
A live run with no source-code edit request continues without an intervention.

\subsection{Plan Generation and Online Steering}
\label{sec:guidance}
\label{sec:online}

\paragraph{Interception and evidence.}
During a live run, \ourmodel{} checks each proposed action before execution.
On the first source-code edit $a_{k^\ast}$, it blocks the edit and
constructs a planning transcript from the completed history $h_{k^\ast-1}$.
The pending response, its edit payload, and its attached reasoning are
excluded.

\begin{algorithm}[!htbp]
\caption{\ourmodel{} in online and offline mode.}
\label{alg:pause}
\small
\begin{minipage}[t]{0.485\linewidth}
\textbf{(a) Online steering}
\begin{algorithmic}[1]
\Require task $(d,R_0)$; Planner $\mathcal{P}$; Executor in harness $H$
\Ensure completed trajectory, or \textsc{Unresolved}
\State $h\gets$ system prompt and $d$; $\mathit{armed}\gets\textbf{true}$
\While{Executor has not finished}
  \State $a\gets$ next proposed action
  \If{$\mathit{armed}$ \textbf{and} $q(a)$}
    \State block $a$; $g\gets\mathcal{P}(d,\phi(h))$
    \If{no usable plan is returned}
      \State \Return \textsc{Unresolved}
    \EndIf
    \State append $g$ to $h$ as a user message
    \State $\mathit{armed}\gets\textbf{false}$
  \Else
    \State $H$ runs $a$; append $a$ and result to $h$
  \EndIf
\EndWhile
\State \Return recorded trajectory
\end{algorithmic}
\end{minipage}\hfill
\begin{minipage}[t]{0.485\linewidth}
\textbf{(b) Offline recovery}
\begin{algorithmic}[1]
\Require task $(d,R_0)$; failed trajectory; Planner $\mathcal{P}$; Executor; budget $B$
\Ensure continued trajectory, or \textsc{Ineligible}
\State $k^\ast\gets$ first source-code edit
\State $c\gets$ last \textsf{think} before $k^\ast$
\If{$k^\ast$ or $c$ does not exist}
  \State \Return \textsc{Ineligible}
\EndIf
\State restore $h_c$, $W_c$, and counter $b_c$
\If{restoration fails the fidelity check}
  \State \Return \textsc{Ineligible}
\EndIf
\State $g\gets\mathcal{P}(d,\phi(h_{c-1}))$
\State append $g$ to $h_c$ as a user message
\State resume Executor from $(h_c,W_c)$ with budget $B-b_c$
\State \Return continued trajectory
\end{algorithmic}
\end{minipage}
\end{algorithm}

\paragraph{Repair plan.}
The Planner $\mathcal{P}$ is a separate language-model invocation with no tools or direct
workspace access. It may use a different underlying model from the Executor
or a separate call to the same model. Given the issue and the transcript,
it identifies the likely root cause, the code that needs to change, and the
intended repair with its rationale:
\begin{equation}
\label{eq:planner}
g=\mathcal{P}\bigl(d,\phi(h_{k^\ast-1})\bigr),
\end{equation}
where $\phi$ renders the recorded evidence as text, omitting editor actions.
The plan $g$ is a
natural-language recommendation; the Executor remains responsible for
implementing and testing it. In Figure~\ref{fig:pause-overview}, the Planner combines
observations from source reads and a reproduction run into a root cause,
the code to change, and the intended repair.

\paragraph{Guided continuation.}
\ourmodel{} appends the plan as a user message to the same Executor's
conversation. The Executor retains the completed history in full, the
workspace $W_{k^\ast-1}$, and its model, system prompt, tools, and decoding
settings. It then chooses its next action with this additional guidance.
The blocked request is not executed automatically: the Executor may revise
its approach or propose another edit to the same file. The interceptor is
disabled after this intervention, and subsequent actions proceed under the
harness's native execution budget.

\subsection{Offline Recovery}
\label{sec:offline}

Offline recovery applies the same planning mechanism to recorded failures.
The first source-code edit locates the intervention, but recovery restores
an earlier reasoning checkpoint rather than pausing a live request. It requires a
trajectory with explicit \textsf{think} actions.

\paragraph{Checkpoint and planning input.}
To place the additional guidance immediately after the Executor's last
recorded reasoning step before the first source-code edit, we select
\begin{equation}
\label{eq:checkpoint}
c=\max\{\,t<k^\ast:a_t\text{ is a }\textsf{think}\,\}.
\end{equation}
The Executor retains that step and its observation in $h_c$, whereas the
Planner receives the preceding transcript and produces
$g=\mathcal{P}(d,\phi(h_{c-1}))$. Thus, the Planner supplies a repair proposal
from the evidence preceding the selected reasoning step, and the Executor
receives both its original reasoning and the new guidance.
Section~\ref{sec:exp-boundary} compares this checkpoint with earlier and
later cuts of the same failures. Recovery retains
the recorded prefix through checkpoint $c$ and generates a new continuation
from $(h_c,W_c)$. A trajectory without a source-code edit request or a preceding
\textsf{think} is ineligible for this recovery protocol.

\paragraph{Restoration and continuation.}
\ourmodel{} rebuilds $h_c$ from the recorded model outputs and observations,
and restores $W_c$ by replaying the prefix actions on $R_0$ in a fresh
container. This retains files created during investigation, such as
reproduction scripts, without calling the Executor model to regenerate
the prefix. It also restores the harness's iteration counter $b_c$;
this counter is distinct from the action index $c$. With an original
budget $B$, continuation receives only $B-b_c$ further iterations.
A fidelity check compares the restored conversation, repository state,
and counter with the recording. Tasks that fail this check are excluded
from the recovery cohort. The plan is then appended using the same message
interface as online steering, and the same Executor resumes with its
retained state and remaining budget.

\paragraph{Use of AI tools.}
Apart from the Executor and Planner models under study, we used AI coding
assistants to implement the harness integrations, write the analysis and
plotting scripts, and assemble the replication package. They produced no
experimental data: every outcome comes from the agent runs and the official
SWE-bench evaluation. The authors reviewed this code, and the released
scripts recompute every reported number from the raw run outputs.

\section{Experiment}
\label{sec:design}
\label{sec:results}

We evaluate the effectiveness and efficiency of \ourmodel{} for online
steering (RQ1) and offline recovery (RQ2), its generalization across harnesses
and reasoning interfaces (RQ3), and the effectiveness of the pre-edit
boundary (RQ4). The shared setup below defines the benchmarks, metrics,
comparisons, and execution protocols used to answer these questions.

\subsection{Experimental Setup}
\label{sec:exp-setup}

\subsubsection{Benchmarks}
\label{sec:exp-benchmarks}

We evaluate \ourmodel{} on SWE-bench Verified~\cite{jimenez2024swebench}, which contains 500 Python
repository-level repair tasks, and SWE-bench Multilingual~\cite{yang2025swesmith}, which contains
300 tasks across nine programming languages. Both benchmarks support the online comparisons
in RQ1, and recorded Verified trajectories support offline recovery in RQ2.
For RQ3, we compare harnesses on Verified. Full-benchmark evaluations measure
overall task resolution, while recovery experiments focus on the Executor's
original failures. The checkpoint comparisons in RQ4 use failures that can
be restored under both conditions.

\subsubsection{Evaluation Metrics}
\label{sec:exp-metrics}

\textit{Repair outcomes.}
We score patches with the corresponding benchmark evaluator.
\emph{Resolution} is the fraction of tasks resolved in the evaluated task
set. For offline resume, tasks that are not continued retain their original
outcomes; recovered failures are added to the original successes and observed
regressions are subtracted.

\textit{Failure recovery and regressions.}
The offline recovery and online model comparisons report
\emph{Recovered}, the number of originally failed tasks solved after
intervention, and \emph{recovery rate}, this count divided by the number of
failures evaluated. Offline recovery uses eligible failures that meet the
checkpoint and restoration requirements in Section~\ref{sec:offline}.
\emph{Regression rate} is the fraction of evaluated baseline successes that
fail after intervention. Changes in these rates are reported in percentage
points.

\textit{Resource use.}
To assess efficiency in RQ1 and RQ2, we report Planner and Executor tokens
separately, in millions, and their model calls where measured.
Dollar costs cover Planner inference at rates of
\$2.50 and \$10 per million input and output tokens, respectively; they do
not include Executor inference. Offline continuation costs exclude the
recorded run used to obtain the prefix. Totals that include this original
run are identified explicitly.

\textit{Scoring and statistical analysis.}
Evaluator errors, empty patches, and missing evaluations count as unresolved.
We compare paired outcomes with exact McNemar tests.

\subsubsection{Baselines and Control Conditions}
\label{sec:baselines}

The unmodified Executor is the baseline.

\textit{Online steering.}
RQ1 compares \ourmodel{} with \textbf{LivePlan}~\cite{liu2026liveplan},
which uses rule-based monitoring to invoke an advisor when it detects trouble,
and the \textbf{SICA overseer}~\cite{robeyns2025sica}, which reviews the
trajectory on a schedule and decides whether to interject. These methods
compare a single pre-edit intervention with rule-triggered advice and
periodic model oversight.

\textit{Offline recovery.}
RQ2 compares \ourmodel{} with \textbf{SHERLOC}~\cite{tamoyan2026sherloc},
which performs a dedicated diagnostic-localization pass before repair and
transfers its findings to the Executor. At a fixed restored checkpoint,
we compare no-plan continuation, self-reconsideration, and Planner guidance
to assess the contribution of the supplied plan. With the same plan, we
compare resuming the recorded state with restarting from the initial
repository to assess the repair--cost trade-off of state reuse.

\textit{Generalization and boundary controls.}
RQ3 compares each harness with its own unmodified Executor. A think-stripped variant removes explicit
\textsf{think} content from the Planner input to test dependence on that
reasoning interface. RQ4 uses guidance from successful donor runs to test
whether failed trajectories can still be redirected at the boundary, then
compares the default offline checkpoint with cuts after 20\% and 90\% of the
recorded actions, an immediate pre-edit cut, and a checkpoint associated
with a later source-code edit.

\subsubsection{Experimental Configuration}
\label{sec:exp-config}

\textit{Models and harnesses.}
We use Qwen3-Coder-30B-A3B-Instruct and Devstral-Small-2507 as Executors.
We hereafter refer to them as Qwen3-coder and Devstral-small, respectively.
The five Planners are GPT-5.6 Sol, Gemini 3.8 Flash, Gemini 3.5 Flash Lite,
Qwen3-30B-A3B-Thinking-2507, hereafter Qwen3-30B-Thinking, and Qwen3-coder.
Online steering runs in OpenHands, SWE-agent~\cite{yang2024sweagent}, and Pi~\cite{zechner2025pi}; offline recovery
uses recorded OpenHands trajectories~\cite{wang2024openhands}.
The main online comparison in RQ1 uses SWE-agent with Qwen3-coder and
GPT-5.6 Sol; RQ3 extends the evaluation to other harnesses.
OpenHands pauses source-code
edit requests through its security and confirmation mechanisms, while SWE-agent and
Pi block the edit tool call directly. OpenHands provides an explicit
\textsf{think} tool; SWE-agent and Pi do not. Planner transcripts include
reasoning text that the Executor writes alongside its actions.

\begin{table}[!htbp]
\centering
\small
\caption{Decoding settings and execution budgets.}
\label{tab:exp-config}
\setlength{\tabcolsep}{4pt}
\renewcommand{\arraystretch}{1.08}
\begin{tabular}{@{}p{0.32\linewidth}p{0.64\linewidth}@{}}
\toprule
\textbf{Component or setting} & \textbf{Configuration} \\
\midrule
Executors & Temperature 0; top-$p$ 1; 4,096 output tokens \\
Executor context & 262,144 tokens (Qwen3-coder); 131,072 (Devstral-small) \\
\ourmodel{} API Planners, default & Temperature 0; 8,192 output tokens \\
Qwen3-30B-Thinking Planner & Temperature 0.99; top-$p$ 0.95; seed 0; 81,920 output tokens \\
LivePlan advisor & Temperature 0; 3,000 output tokens \\
SICA overseer & Temperature 0; 8,192 output tokens \\
\midrule
Recorded Verified runs & 100 iterations \\
SWE-agent comparisons & 100 Executor calls; 14,400 seconds per run  \\
\bottomrule
\end{tabular}
\end{table}

\textit{Planner configuration.}
Table~\ref{tab:exp-config} summarizes decoding and budget settings. Online, the Planner is queried
once when triggered; no usable plan within the output limit counts as
unresolved. Offline, malformed Planner output is retried before inserting
one plan.

\textit{Evaluation protocol.}
The main comparisons use one designated attempt per task. The donor-guidance
diagnostic in RQ4 uses three attempts per donor--recipient pair and counts
a pair as successful if at least one continuation succeeds. Online attempts
interrupted by an endpoint connection, container, or runtime-start failure
are rerun.

In RQ2, resumed runs use the recorded remaining Executor budget, whereas
restarted runs receive the full budget. In the paired checkpoint comparisons
of RQ4, each arm independently restores its prefix, obtains a plan from that
prefix, and receives the same number of fresh Executor iterations. The cuts
after 20\% and 90\% of the actions are placed by the length of the recorded
failed trajectory.

\subsection{RQ1: How Effective and Efficient Is Online Steering?}
\label{sec:exp-overall}

The central design choice in \ourmodel{} is to provide one repair plan when
an Executor first requests a source-code edit. We first evaluate this choice
during live execution: does a single intervention improve task resolution,
and how much computation does it require compared with rule-based monitoring
and periodic model oversight? This establishes the effectiveness of online
steering before considering recovery from recorded failures.

\textit{Comparison setup.}
Table~\ref{tab:comparison} reports results in SWE-agent with Qwen3-coder
as the Executor and GPT-5.6 Sol as the Planner. We compare \ourmodel{} with
LivePlan and SICA on all 500 Verified and 300 Multilingual tasks, using
matched Executor budgets, together with the unmodified Executor on each benchmark. We compare
resolution alongside Planner and Executor tokens; dollar costs cover
Planner inference.

\begin{table*}[t]
\centering
\small
\caption{\textbf{Steering compared with monitoring-based methods.}
Qwen3-coder Executor and GPT-5.6 Sol Planner in SWE-agent on all 500
SWE-bench Verified and all 300 SWE-bench Multilingual tasks.
\emph{Resolution}: resolved tasks over the full benchmark, with the gain
over each block's baseline. Tokens are in millions; \emph{Cost} covers
Planner inference only.}
\label{tab:comparison}

\setlength{\tabcolsep}{6pt}
\renewcommand{\arraystretch}{1.08}

\begin{tabular}{@{}llccccc@{}}
\toprule
\multicolumn{1}{@{}l}{\multirow{2}{*}{\textbf{Method}}} &
\multicolumn{1}{l}{\multirow{2}{*}{\textbf{Planner}}} &
\multicolumn{1}{l}{\multirow{2}{*}{\textbf{Resolved}}} &
\multicolumn{1}{l}{\multirow{2}{*}{\textbf{Resolution (\%)}}} &
\multicolumn{2}{c}{\textbf{Tokens (M)}} &
\multicolumn{1}{l@{}}{\multirow{2}{*}{\textbf{Cost (\$)}}} \\
\cmidrule(lr){5-6}
& & & & \multicolumn{1}{l}{\textbf{Planner}} & \multicolumn{1}{l}{\textbf{Executor}} & \\
\midrule
\multicolumn{7}{@{}l}{\textit{SWE-bench Verified, 500 tasks}} \\
Baseline & -- & 251/500 & 50.2 & 0 & 595.5 & 0 \\
LivePlan~\cite{liu2026liveplan} & GPT-5.6 Sol
& 309/500 & 61.8 {\scriptsize (\gain{+11.6})} & 9.2 & 561.9 & 26.10 \\
SICA overseer~\cite{robeyns2025sica} & GPT-5.6 Sol
& 338/500 & 67.6 {\scriptsize (\gain{+17.4})} & 80.9 & 1056.4 & 231.71 \\
\textbf{\ourmodel{}} & GPT-5.6 Sol
& \textbf{325/500} & \textbf{65.0 {\scriptsize (\gain{+14.8})}}
& 9.0 & 564.2 & 26.13 \\

\midrule
\multicolumn{7}{@{}l}{\textit{SWE-bench Multilingual, 300 tasks}} \\
Baseline & -- & 85/300 & 28.3 & 0 & 469.3 & 0 \\
LivePlan~\cite{liu2026liveplan} & GPT-5.6 Sol
& 118/300 & 39.3 {\scriptsize (\gain{+11.0})} & 5.9 & 497.5 & 16.41 \\
SICA overseer~\cite{robeyns2025sica} & GPT-5.6 Sol
& 133/300 & 44.3 {\scriptsize (\gain{+16.0})} & 55.9 & 1003.5 & 159.63 \\
\textbf{\ourmodel{}} & GPT-5.6 Sol
& \textbf{129/300} & \textbf{43.0 {\scriptsize (\gain{+14.7})}} & 8.5 & 459.9 & 23.68 \\

\bottomrule
\end{tabular}
\end{table*}

\begin{figure*}[t]
\centering
\begin{minipage}[t]{0.48\linewidth}
  \centering
  \includegraphics[width=\linewidth]{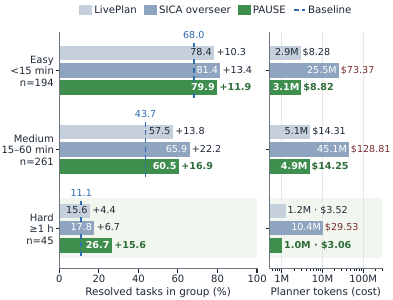}
\caption{Resolution and Planner cost by task difficulty on SWE-bench
Verified with SWE-agent. Dashed lines show the unmodified Executor, and
the accompanying values show percentage-point gains over it. Here,
$n$ denotes the number of tasks in each difficulty group.}
  \label{fig:difficulty-methods}
\end{minipage}\hfill
\begin{minipage}[t]{0.48\linewidth}
  \centering
  \includegraphics[width=\linewidth]{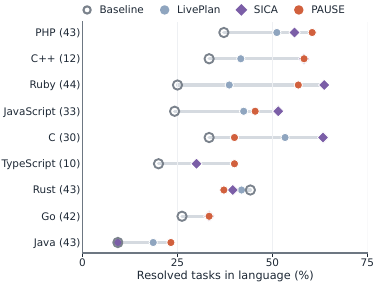}
\caption{Resolution by programming language on SWE-bench Multilingual
with SWE-agent. The bar in each row spans the lowest and highest
resolution among the four configurations, and the number after each
language is its task count.}
  \label{fig:language-methods}
\end{minipage}
\end{figure*}

\textit{Overall repair gains.}
\ourmodel{} improves resolution on both benchmarks while reducing Executor token use. On
Verified, resolution rises from 50.2\% to 65.0\%, a gain of 14.8 percentage
points and a net increase of 74 solved tasks. On Multilingual, it rises
from 28.3\% to 43.0\%, adding 14.7 points and 44 solved tasks, a difference
that is statistically significant ($p=4\times10^{-7}$). Executor tokens decrease from
595.5M to 564.2M on Verified and from 469.3M to 459.9M on Multilingual,
reductions of 5.3\% and 2.0\%, respectively, while the intervention adds 9.0M
and 8.5M Planner tokens. These
results show that the repair gains coincide with lower Executor token use,
with the additional computation supplied by the Planner.

\textit{Comparison with rule-based steering.}
Against LivePlan, \ourmodel{} achieves higher observed resolution at
similar Planner cost, a gain of 3.2 percentage points that is not
statistically significant ($p=0.11$). \ourmodel{} makes 0.93 Planner calls per task
against LivePlan's 2.33, yet their aggregate Planner tokens are close,
9.0M versus 9.2M, as are their costs, \$26.13 versus \$26.10. Thus, fewer
planning requests here correspond to a similar total planning expenditure.
On Multilingual, \ourmodel{} leads LivePlan by 3.7 percentage points,
129 against 118 solved tasks, a difference that is also not statistically
significant ($p=0.18$). Its Planner cost remains similar, \$23.68 versus
\$16.41, while it uses 7.6\% fewer Executor tokens, 459.9M versus 497.5M.
These results support intervention at the first source-code edit as an effective
alternative to rule-triggered advice in the evaluated settings.

\textit{Comparison with periodic oversight.}
SICA achieves the highest aggregate resolution at 67.6\%,
2.6 percentage points and 13 tasks above \ourmodel{}'s 65.0\%; this
difference is not statistically significant ($p=0.18$).
SICA's resource use is substantially higher: it requires 21 times as many Planner calls, 8.9 times the Planner
cost, and 1.9 times the Executor tokens; the Planner expenditure is
\$231.71 versus \$26.13, and Executor token use is 1,056.4M versus 564.2M.
Periodic oversight therefore does not yield significantly higher resolution
for its additional computation: \ourmodel{} stays within a non-significant
margin of SICA at about 11\% of its Planner cost.
The same pattern holds on Multilingual, where SICA resolves 133 tasks
against 129 for \ourmodel{}, a lead of 1.3 percentage points that is not
statistically significant ($p=0.71$). SICA again incurs 6.7 times the
Planner cost, \$159.63 versus \$23.68, and 2.2 times the Executor tokens,
1,003.5M versus 459.9M, so \ourmodel{} stays within this margin at about
15\% of SICA's Planner cost.
A single intervention can thus provide repair benefits
comparable to repeated oversight with far less planning and execution.

\textit{Task difficulty.}
Figure~\ref{fig:difficulty-methods} shows that \ourmodel{} improves over the
unmodified Executor and exceeds LivePlan in all three difficulty groups,
while its ranking relative to SICA varies. SICA leads by 1.5 percentage points on easy tasks and 5.4 on medium tasks,
but at 8.3 and 9.0 times the Planner cost of \ourmodel{}, spending \$73.37
against \$8.82 and \$128.81 against \$14.25, respectively. These groups contain 194 and
261 tasks, respectively, together accounting for 91\% of Verified; their
weight explains SICA's higher aggregate resolution on Verified. In the 45-task hard
group, \ourmodel{} raises resolution from the unmodified Executor's 11.1\%
to 26.7\%, exceeding SICA's 17.8\% by 8.9 points, with Planner cost of
\$3.06 against \$29.53. The hard-task gains suggest that external planning
may help the Executor turn its accumulated investigation into a more
effective repair strategy. By intervening before the first source-code
edit, \ourmodel{} gives the Planner an opportunity to reassess how the
collected evidence should translate into a concrete change. A repair plan
grounded in the accumulated evidence may therefore be useful even when
the Executor's unaided success rate is low.

\textit{Programming languages.}
Figure~\ref{fig:language-methods} shows that \ourmodel{} improves over the
unmodified Executor in eight of the nine Multilingual languages. The largest
gains come from Ruby, PHP, and JavaScript, which add 14, 10, and 7 solved
tasks, respectively, and together account for 31 of the 44 additional tasks.
Rust is the only language in which \ourmodel{} falls below the unmodified
Executor, resolving 16 of 43 tasks against 19. LivePlan and SICA also fall
below it on Rust, with 18 and 17 tasks, so the drop is shared by all three
forms of external guidance rather than specific to intervention at the first
source-code edit. SICA's lead on Multilingual comes mainly from C and Ruby,
where it solves 7 and 3 more tasks than \ourmodel{}, while \ourmodel{} leads
SICA on Java, 10 against 4. C++ and TypeScript contain only 12 and 10 tasks,
so their per-language differences are not informative.

\rqanswer{With SWE-agent, \ourmodel{} raises resolution
by 14.8 percentage points on Verified and 14.7 on Multilingual while reducing Executor token use. It
achieves higher observed resolution than LivePlan on both benchmarks at similar Planner cost.
Against SICA, it trails by a non-significant 2.6 points on Verified and 1.3
on Multilingual at about 11\% and 15\% of
SICA's Planner cost, and it leads on hard Verified tasks.}

\subsection{RQ2: How Effective and Efficient Is Offline Recovery?}
\label{sec:exp-offline}
\label{sec:exp-insert}
\label{sec:exp-cost}

Beyond steering a live run, \ourmodel{} can revisit a failed attempt by
resuming from its recorded pre-edit prefix. This setting allows the agent
to reuse its repository investigation while following a new repair plan.
We evaluate whether such guidance recovers failures and whether retaining
the prefix reduces the computation needed compared with starting again.

\textit{Comparison setup.}
We use recorded OpenHands trajectories on SWE-bench Verified.
Table~\ref{tab:main} compares three continuation conditions on 223 eligible
Qwen3-coder failures: \textbf{no plan (N)}, which resumes without an added
message; \textbf{self-reconsideration (S)}, which flags the selected
\textsf{think} as a possible cause of failure and asks the Executor to
reconsider; and \textbf{Planner plan (P)}, which supplies a repair plan from
the task and pre-edit transcript. All three retain the same checkpoint,
conversation history, workspace, and remaining Executor budget. We evaluate
four Planners with each Executor, including 174 eligible Devstral-small
failures. The table reports recovery among these failures.
Prior analyses of LLM self-correction emphasize distinguishing
the sources of feedback when evaluating revision~\cite{huang2024large,kamoi2024can}.

\begin{table*}[!htbp]
\centering
\small
\caption{\textbf{Offline recovery in OpenHands on SWE-bench Verified across Executors.}
Each failed trajectory resumes from its restored checkpoint, the last
\textsf{think} before the first source-code edit, with its remaining budget.
Recovered reports solved eligible failures: 223 for Qwen3-coder and 174 for
Devstral-small. N: no plan; S: self-reconsideration; P: plan from the named Planner.}
\label{tab:main}
\setlength{\tabcolsep}{7pt}
\renewcommand{\arraystretch}{1.08}

\begin{tabular}{@{}lcc@{}}
\toprule
\textbf{Intervention}
& \textbf{Qwen3-coder Executor}
& \textbf{Devstral-small Executor} \\
\midrule
N: no plan & 9/223 (4.0\%) & 14/174 (8.0\%) \\
S: self-reconsideration & 13/223 (5.8\%) & 17/174 (9.8\%) \\
\addlinespace[2pt]
P: Executor Model & 12/223 (5.4\%) & 20/174 (11.5\%) \\
P: Gemini 3.5 Flash Lite & 22/223 (9.9\%) & 12/174 (6.9\%) \\
P: Gemini 3.8 Flash & 31/223 (13.9\%) & 29/174 (16.7\%) \\
\addlinespace[2pt]
\textbf{P: GPT-5.6 Sol} & \textbf{79/223 (35.4\%)} & \textbf{53/174 (30.5\%)} \\
\bottomrule
\end{tabular}
\end{table*}

\textit{Recovery through external guidance.}
The continuation controls distinguish another opportunity to execute from
the contribution of a new repair plan. Unaided continuation recovers 9 of
223 failures (4.0\%), showing that some failed attempts can succeed simply
by continuing from the restored checkpoint. Self-reconsideration recovers
13 (5.8\%), so a cue to revisit the selected reasoning yields limited
recovery in this evaluation. With GPT-5.6 Sol, \ourmodel{} recovers 79
failures (35.4\%). These
conditions share the restored state and remaining Executor budget,
supporting the value of external repair guidance. One explanation
is that spotting a flawed line of reasoning still leaves
the Executor to find a replacement, whereas the Planner
supplies a concrete direction.

\textit{The choice of Planner matters.}
Using the Executor's own model as a separate Planner recovers 12 failures
(5.4\%), close to the observed counts for unaided continuation and
self-reconsideration. Introducing a planning role therefore does not by
itself ensure a large recovery gain. Gemini 3.8 Flash recovers 31 failures,
while GPT-5.6 Sol recovers 79. These differences
suggest that the benefit depends on the guidance supplied by the chosen
Planner, beyond the presence of an additional model call. With Devstral-small, GPT-5.6 Sol recovers 53 of 174 failures
(30.5\%), showing that this
recovery benefit also extends to a second Executor.

\textit{Full-benchmark comparison.}
Table~\ref{tab:offline-comparison} reports the separate 500-task evaluation
with Qwen3-coder, comparing \ourmodel{} with SHERLOC under
Qwen3-30B-Thinking planning and additionally evaluating GPT-5.6 Sol.
\textbf{Resume} retains the recorded history and workspace and uses the
remaining budget; \textbf{restart} gives the same plan to a fresh Executor
with the initial repository and full budget. The table reports net
resolution, accounting for observed regressions, alongside Planner and
Executor tokens. It therefore complements the failure-recovery results
with overall repair outcomes and resource use.

\begin{table}[!htbp]
\centering
\small
\caption{\textbf{Full-benchmark offline evaluation on SWE-bench Verified.}
Qwen3-coder in OpenHands, 500 tasks. Resume retains history and
workspace with the remaining budget; restart uses the same plan from the
initial repository with the full budget. Resume counts add recoveries
to original successes and subtract observed regressions. Tokens (M)
exclude the original run for continuations; costs cover Planner inference.}
\label{tab:offline-comparison}
\setlength{\tabcolsep}{4pt}
\renewcommand{\arraystretch}{1.08}
\begin{tabular}{@{}llccccc@{}}
\toprule
\multirow{2}{*}{\textbf{Method}} & \multirow{2}{*}{\textbf{Planner}} & \multirow{2}{*}{\textbf{Resolved}} & \multirow{2}{*}{\textbf{Resolution (\%)}} &
\multicolumn{2}{c}{\textbf{Tokens (M)}} & \multirow{2}{*}{\textbf{Cost (\$)}} \\
\cmidrule(lr){5-6}
& & & & \textbf{Planner} & \textbf{Executor} & \\
\midrule
Baseline & -- & 228/500 & 45.6 & 0 & 791.9 & 0 \\
SHERLOC~\cite{tamoyan2026sherloc} & Qwen3-30B-Thinking
& 211/500 & 42.2 {\scriptsize ($-$3.4)} & 65.7 & 703.9 & 0 \\
\ourmodel{} (restart) & Qwen3-30B-Thinking
& 207/500 & 41.4 {\scriptsize ($-$4.2)} & 16.0 & 641.6 & 0 \\
\ourmodel{} (resume) & Qwen3-30B-Thinking
& 217/500 & 43.4 {\scriptsize ($-$2.2)} & 14.5 & 375.7 & 0 \\
\ourmodel{} (restart) & GPT-5.6 Sol
& 293/500 & 58.6 {\scriptsize (\gain{+13.0})} & 8.7 & 519.1 & 25.75 \\
\textbf{\ourmodel{} (resume)} & GPT-5.6 Sol
& \textbf{291/500} & \textbf{58.2 {\scriptsize (\gain{+12.6})}}
& 8.7 & \textbf{347.4} & 25.75 \\

\bottomrule
\end{tabular}
\end{table}

\textit{Reusing a failed investigation.}
The resume--restart comparison tests whether a failed attempt can still
provide a useful starting point for another repair. It measures the
practical trade-off of prefix reuse, with both retained state and available
budget changing. On the 223 failures, the same GPT-5.6 Sol plans recover
79 tasks with resume and 81 with restart. Restart requires 2.0$\times$ as
many Executor calls, or 6,020 additional calls, and 22\% more Executor
tokens, or 56.1M. These comparisons show that a failed run can still provide a useful
starting state for recovery, with a trade-off between execution cost and
the number of tasks recovered.

At full-benchmark scale, GPT-5.6 Sol resume achieves 58.2\% resolution
against restart's 58.6\%, saving 171.7M Executor tokens (33\%) with the
same 8.7M Planner tokens. With Qwen3-30B-Thinking, resume saves 265.9M
Executor tokens (41\%) and resolves more tasks than restart.

\textit{Cost of the recorded investigation.}
The savings above concern continuation from an existing trajectory.
Relative to SHERLOC, resume with Qwen3-30B-Thinking uses 47\% fewer Executor
tokens and 78\% fewer Planner tokens; GPT-5.6 Sol resume uses approximately
half as many Executor tokens as SHERLOC. Including the original recorded
run, however, GPT-5.6 Sol resume uses 1,139.3M Executor tokens against
SHERLOC's 703.9M. The practical use case is therefore recovery after an
agent has already attempted the task: the recorded investigation provides
an existing source of evidence and execution state, allowing another
repair attempt at lower additional cost.

\textit{Outcomes on originally solved tasks.}
An intervention that recovers failures could also break repairs that the
Executor originally completed. We therefore examine the 178 originally
solved tasks evaluated under both Planners; of the other 50 original
successes, 13 never request a source-code edit, so \ourmodel{} leaves them
unchanged, and 37 could not be successfully replayed in a fresh container in at
least one Planner's run. GPT-5.6 Sol plans lead to 10 regressions, a
regression rate of 5.6\%, whereas Qwen3-30B-Thinking plans lead to 26, a
rate of 14.6\%; the difference between the two Planners is significant
at $p<0.01$. To determine how many of these regressions arise from
resumption itself, we resume the same checkpoints with no plan. This
control yields 10 regressions, the same number as under GPT-5.6 Sol, and the
difference between the two is not statistically significant at $p=0.81$.
Restoring a checkpoint and re-executing the remaining steps therefore
accounts for most of the observed regressions, and guidance from a strong
Planner does not increase them. Plans from Qwen3-30B-Thinking, by contrast,
produce 16 more regressions than the no-plan control, suggesting that weaker
guidance can steer the Executor away from repairs it would otherwise
complete. Because the full-benchmark resolution subtracts every observed
regression, including those that resumption alone produces, it provides a
conservative estimate of the gain attributable to the plan. Together with
the recovery results, these findings show that GPT-5.6 Sol guidance
recovers failed runs without breaking more successful runs than
continuation alone.

\textit{Net repair performance.}
The full-benchmark results establish whether failure recovery translates
into an overall improvement after accounting for observed regressions.
Resuming with GPT-5.6 Sol reaches 58.2\% resolution, 12.6 percentage points
above the unmodified Executor's 45.6\%. With Qwen3-30B-Thinking, resume reaches 43.4\%, compared with SHERLOC's
42.2\%; both remain below the unmodified Executor. The resource savings
with this Planner therefore accompany a loss in absolute repair performance.
GPT-5.6 Sol raises resolution by 16.0 points over SHERLOC, although this
comparison also changes the Planner. Taken together, the results show why
recovery quality and reuse efficiency matter jointly: a capable Planner
can turn retained investigation into improved repair outcomes, while prefix
reuse reduces the execution required for another attempt.

\rqanswer{Offline \ourmodel{} can make productive use of a failed
investigation by supplying new repair guidance. With GPT-5.6 Sol, it
recovers 35.4\% of eligible Qwen3-coder failures and 30.5\% of
Devstral-small failures; the separate full-benchmark evaluation raises net
resolution from 45.6\% to 58.2\%, and its plans cause no more regressions on
originally solved tasks than resuming without a plan. Given an existing
trajectory, resuming
uses 33--41\% fewer Executor tokens than restarting at full-benchmark scale.
Substantial recovery depends on the Planner, while prefix reuse offers a
practical trade-off \mbox{between additional execution cost and repair outcomes.}}

\subsection{RQ3: Does \ourmodel{} Generalize Across Settings?}
\label{sec:exp-generalization}

We next examine whether the benefits of online steering extend across
execution harnesses and reasoning interfaces, which vary across coding
agents and determine the actions and transcript that \ourmodel{} uses.

\textit{Comparison setup.}
Table~\ref{tab:harness-results} compares SWE-agent, Pi, and OpenHands on
all 500 Verified tasks with the Qwen3-coder Executor and GPT-5.6 Sol
Planner fixed. Because the harnesses differ in tools, system prompts, and
budgets, each is compared with its own unmodified Executor. Only OpenHands
exposes an explicit \textsf{think} tool, and we also remove this content
from its Planner input.

\begin{table}[!htbp]
\centering
\small
\caption{\textbf{Transfer across agent harnesses.} Online \ourmodel{} with Qwen3-coder Executor and GPT-5.6 Sol Planner on 500 Verified tasks. \emph{Think stripped} hides \textsf{think} content from the Planner; units follow Table~\ref{tab:comparison}.}
\label{tab:harness-results}

\setlength{\tabcolsep}{6pt}
\renewcommand{\arraystretch}{1.08}

\begin{tabular}{@{}lllcccc@{}}
\toprule
\multirow{2}{*}{\textbf{Harness}} & \multirow{2}{*}{\textbf{Method}} & \multirow{2}{*}{\textbf{Resolved}} & \multirow{2}{*}{\textbf{Resolution (\%)}} &
\multicolumn{2}{c}{\textbf{Tokens (M)}} & \multirow{2}{*}{\textbf{Cost (\$)}} \\
\cmidrule(lr){5-6}
& & & & \textbf{Planner} & \textbf{Executor} & \\
\midrule

\multirow{2}{*}{\shortstack[l]{SWE-agent\\{\scriptsize no \textsf{think} tool}}}
& Baseline & 251/500 & 50.2 & 0 & 595.5 & 0 \\
& \textbf{\ourmodel{}} & \textbf{325/500} & \textbf{65.0 {\scriptsize (\gain{+14.8})}} & 9.0 & 564.2 & 26.13 \\
\midrule

\multirow{2}{*}{\shortstack[l]{Pi\\{\scriptsize no \textsf{think} tool}}}
& Baseline & 185/500 & 37.0 & 0 & 341.9 & 0 \\
& \textbf{\ourmodel{}} & \textbf{215/500} & \textbf{43.0 {\scriptsize (\gain{+6.0})}} & 6.7 & 358.8 & 18.57 \\
\midrule

\multirow{3}{*}{\shortstack[l]{OpenHands\\{\scriptsize \textsf{think} tool}}}
& Baseline & 227/500 & 45.4 & 0 & 372.5 & 0 \\
& \textbf{\ourmodel{}} & \textbf{311/500} & \textbf{62.2 {\scriptsize (\gain{+16.8})}} & 6.7 & 390.6 & 20.38 \\
&  think stripped & 318/500 & 63.6 {\scriptsize (\gain{+18.2})} & 6.1 & 371.7 & 18.79 \\

\bottomrule
\end{tabular}
\end{table}

\textit{Across harnesses.}
\ourmodel{} improves resolution in all three harnesses, by 14.8 percentage
points in SWE-agent, 6.0 in Pi, and 16.8 in OpenHands, or 74, 30, and 84
additional solved tasks. In OpenHands, it solves 106 tasks that the
unmodified Executor misses while losing 22 ($p=2\times10^{-14}$). The gains
vary even with the model pair fixed: the plan narrows the gap between
SWE-agent and OpenHands from 4.8 to 2.8 points, whereas Pi gains least
despite the lowest baseline. One likely contributor is that the Pi Executor
sometimes modified source code through shell commands before its first edit
tool call, so no plan was inserted (Section~\ref{sec:limitations}); we did
not measure how often this occurs.

\textit{Cost of transfer.}
The Planner cost stays between \$0.037 and \$0.052 per task, and Planner
tokens remain below 2\% of the Executor tokens in every harness. Executor tokens fall by 5.3\% in SWE-agent but rise by 4.9\% in both OpenHands and Pi, so the Executor savings observed in RQ1 depend on how the harness
continues after receiving the plan.

\textit{Where the online gain arises.}
Splitting the 500 OpenHands tasks by the outcome of the recorded
trajectories used in RQ2 shows gains in every stratum. Resolution rises
from 20.6\% to 40.4\% on the 223 tasks of the offline recovery cohort,
from 16.3\% to 44.9\% on the 49 failures excluded from offline replay, and
from 75.9\% to 87.3\% on the 228 recorded successes, where \ourmodel{}
gains 39 tasks and loses 13. The plan thus reaches failures that offline
recovery cannot restore and also raises success on tasks within the
Executor's reach.

\textit{Reasoning interfaces.}
Removing \textsf{think} content from the OpenHands Planner input yields 318
solved tasks against 311 with the full transcript, at nearly the same cost.
The two variants differ on 71 tasks, split 39 to 32 ($p=0.48$), so explicit
reasoning changes which tasks a plan helps but not significantly how many. Together with
the gains in SWE-agent and Pi, which have no \textsf{think} tool, this
indicates that reads, commands, and their outputs give the Planner
sufficient evidence for a useful plan.

\rqanswer{Online \ourmodel{} raises resolution by 6.0--16.8 percentage
points across three harnesses on Verified at a Planner cost of at most
\$0.052 per task. In OpenHands, it improves resolution on both recorded
failures and recorded successes. Gains persist without explicit
\textsf{think} content, while their magnitude and the effect on Executor
tokens vary with the harness.}

\subsection{RQ4: Is the Pre-Edit Boundary Effective?}
\label{sec:exp-boundary}
The preceding evaluations assess what \ourmodel{} achieves; we now examine
the intervention point underlying those results. The first source-code edit
occurs at a median of 60.0\% of action progress on Verified
(Figure~\ref{fig:checkpoint-position-distribution}), after substantial
investigation. This raises two questions: can guidance still redirect the
repair at this point, and would an earlier or later point improve
recovery? We examine both through interventions on recorded trajectories.

\textit{Comparison setup.}
We first insert pre-edit reasoning from a successful run of the same task
as donor guidance, testing whether a failed trajectory can produce a
successful continuation. We then compare the default offline checkpoint,
after the last \textsf{think} before the first source-code edit, with three
alternative cuts of the same recorded failures: after 20\% of the recorded
actions, immediately before the first source-code edit, and after 90\% of
the recorded actions. We also compare it with a checkpoint associated with a later source-code edit. Each arm independently
restores its prefix and obtains its own GPT-5.6 Sol plan, with the same number of fresh Executor iterations. Each comparison thus tests the
checkpoint together with its plan.

\textit{Guidance at the boundary.}
Donor guidance produces a successful continuation in at least one of three
attempts for 64 of 97 donor--recipient pairs, a rate of 66.0\%, and in all
three attempts for 41, a rate of 42.3\%. Donor and recipient runs come from
Executor configurations with different prompts, and across the four
donor--recipient combinations the rate of at least one success ranges from
58.8\% to 69.0\%. This diagnostic uses
guidance drawn from successful executions, so it assesses the possibility
of a successful continuation at the boundary. The retained investigation
remains compatible with a correct repair even when the original run failed.

\textit{Earlier and later cuts.}
Restoring a prefix replays its recorded actions in a fresh container, and
this replay can fail the fidelity check or leave a continuation without a
grading result. A task enters a comparison only when every arm in that
comparison restores and is graded. The four-way comparison therefore uses
the 161 tasks graded in all arms, and each pairwise comparison with
\ourmodel{} uses the larger set shared by that pair, which contains 207,
196, or 167 tasks.
Table~\ref{tab:pause-position} compares the four pause positions. On the
161 tasks graded in every arm, the \ourmodel{} checkpoint recovers 56 tasks,
a rate of 34.8\%, compared with 43 for the cut after 20\% of the actions
and 50 for both the cut immediately before the first source-code edit and
the cut after 90\%. In every paired comparison, \ourmodel{} solves more
tasks that the alternative misses than the reverse. The early cut performs
worst: on 207 shared tasks, \ourmodel{} solves 26 tasks that the 20\% cut
misses, while the 20\% cut solves 15 that \ourmodel{} misses, a gap of
5.3 percentage points. This difference is a trend that is not significant
at the 0.05 level ($p=0.12$), but its direction is consistent with a
Planner that has seen less of the investigation. Cutting immediately
before the edit retains the intervening actions but recovers 59 of 196
shared tasks, compared with 65 for \ourmodel{}. The cut after 90\% lets
the Planner see almost the entire failed run, by which point the Executor
has already edited, yet it recovers 32.3\% of 167 shared tasks, compared
with 34.7\% for \ourmodel{}. Neither later cut differs significantly from
\ourmodel{} ($p=0.39$ and $p=0.62$). Recovery is therefore highest at the
checkpoint associated with the first source-code edit, which \ourmodel{}
locates without knowing the length of the run.

\begin{table}[!htbp]
\centering
\small
\caption{\textbf{Recovery under different pause positions on SWE-bench Verified.}
Each arm restores its own prefix of the same recorded failures and obtains
its own GPT-5.6 Sol plan. \emph{Common}: the 161 tasks graded in every arm.
\emph{Paired}: each arm against the \ourmodel{} checkpoint on their largest
shared set, with the tasks solved only by the arm or only by \ourmodel{}
and the exact McNemar $p$-value.}
\label{tab:pause-position}
\setlength{\tabcolsep}{4pt}
\renewcommand{\arraystretch}{1.08}
\begin{tabular}{@{}lcccccc@{}}
\toprule
\multirow{2}{*}{\textbf{Pause position}} & \textbf{Common} & \multicolumn{5}{c}{\textbf{Paired with \ourmodel{}}} \\
\cmidrule(lr){2-2}\cmidrule(lr){3-7}
& \textbf{Resolved (\%)} & \textbf{Tasks} & \textbf{Arm (\%)} & \textbf{\ourmodel{} (\%)} & \textbf{Only arm / \ourmodel{}} & $p$ \\
\midrule
\textbf{\ourmodel{} checkpoint} & \textbf{56/161 (34.8)} & -- & -- & -- & -- & -- \\
\midrule
After 20\% of actions & 43/161 (26.7) & 207 & 27.5 & 32.9 & 15 / 26 & 0.12 \\
Immediately before the edit & 50/161 (31.1) & 196 & 30.1 & 33.2 & 14 / 20 & 0.39 \\
After 90\% of actions & 50/161 (31.1) & 167 & 32.3 & 34.7 & 16 / 20 & 0.62 \\
\bottomrule
\end{tabular}
\end{table}

\textit{Waiting for a later source-code edit.}
As a supplementary check on a smaller cohort, we tie the later cut to the
Executor's own actions rather than to a fixed fraction of the run, selecting
the first later source-code edit preceded by a different reasoning step. Of 46 candidate trajectories, 17 fail restoration checks
before any model call, leaving 29 paired runs. The first and later source-code edit
checkpoints each recover 15 tasks. Consistent with the cut after 90\%,
waiting for another source-code edit yields no increase in recovered tasks.

\rqanswer{Successful continuations remain possible at the pre-edit
boundary: donor guidance yields success for 64 of 97
pairs within three attempts. Under matched continuation budgets, no
alternative pause position, whether after 20\% or 90\% of the actions,
immediately before the first source-code edit, or at a later source-code
edit, recovers more tasks than the default offline checkpoint, and an early
cut shows a trend toward fewer recoveries. These
results support the first source-code edit as a practical point for steering
after investigation and before the requested edit executes.}

\section{Limitations and Threats to Validity}
\label{sec:threats}

\subsection{Limitations}
\label{sec:limitations}

\ourmodel{} relies on repository evidence accumulated before the
intervention. The Planner receives the issue and the pre-edit transcript,
but cannot independently inspect the workspace or execute commands. When
the investigation omits relevant code or informative reproduction feedback,
the Planner can propose hypotheses but cannot verify them against the
repository. Prior studies distinguish correction based only on
model-generated feedback from correction supported by reliable external
feedback~\cite{huang2024large,kamoi2024can}.
Further investigation by the Executor may therefore be needed
to assess whether the suggested strategy is appropriate.

The framework supplies a single repair plan and does not refresh its
guidance as new observations arrive. New test results or implementation
discoveries may make parts of that plan obsolete, leaving subsequent
adjustment to the Executor. Steering can redirect the repair strategy,
but translating that strategy into a correct patch still requires the
Executor to implement and validate the changes.

The intervention depends on a file-editor request for the first source-code edit.
Integrating \ourmodel{} requires a harness that exposes this request
before execution and a path rule appropriate to the repository layout.
Trajectories without such a request receive no intervention, limiting
coverage in workflows whose source changes use other editing interfaces,
such as shell commands.

Offline recovery further requires a recorded trajectory with a usable
reasoning checkpoint and faithfully restorable history and workspace.
A transcript alone is insufficient when the corresponding execution state
cannot be reconstructed. The savings from recovery
concern the additional Executor computation for another attempt once the
prefix is available; producing the original trajectory remains part of
the total cost. Online steering uses the live state without prior recording.

\subsection{Threats to Validity}
\label{sec:validity}

\subsubsection{Internal validity.}
SWE-bench Verified and SWE-bench Multilingual issues and fixes are public and may have appeared in a
Planner's training data. We checked agent inputs, Planner prompts,
repository snapshots, and container network access for direct exposure
during execution; these checks cannot determine prior training exposure.
All Executor runs mount only the instance ID and problem statement. The offline regression comparisons use one run per task, so small
differences may be noise.

\subsubsection{External validity.}
The evaluation covers SWE-bench Verified, SWE-bench Multilingual, and the tested Planner,
Executor, and harness configurations. Their diversity provides evidence
across several settings, but transfer to substantially different
repositories or longer, multi-stage development workflows remains
unexamined. Beyond Python, the evidence comes from the online SWE-agent
comparison on Multilingual; offline recovery and the harness comparison
use Verified only. The pause-position ablation covers offline recovery with
Qwen3-coder on Verified only. Offline recovery results additionally concern trajectories
that meet the checkpoint and restoration requirements; failures outside
this eligible cohort may respond differently to guidance.

\subsubsection{Construct validity.}
The first source-code edit operationalizes the intervention point through an
action type and a benchmark-specific path rule. Broader rules can admit reproduction scripts,
so the detected event need not represent the same stage of investigation
in every repository. The alternative cuts at 20\% and 90\% of the actions
are defined by the length of the recorded run, which is unknown during a
live run, so they serve as offline reference points rather than deployable
triggers. Benchmark tests provide the measure of repair success;
behavior beyond their coverage remains unevaluated. Incomplete grading
records count as unresolved.

\subsubsection{Conclusion validity.}
The main comparisons use one attempt per task, while the donor-guidance
diagnostic uses three attempts per pair. The paired pause-position
comparisons in RQ4 use 167 to 207 tasks, and none of them reaches
significance at the 0.05 level; the lower recovery of the early cut is a
trend that repeated runs would need to confirm. Similar observed
recovery counts do not establish equivalence; repeated paired runs are
needed to assess whether small differences persist.

\section{Future Work}
\label{sec:future-work}

The results motivate two directions for studying and extending
the benefits of a pre-edit plan.

\paragraph{Evidence selection for planning.}
Removing explicit reasoning produces similar aggregate resolution but
different task outcomes, leaving open which parts of an investigation
support effective guidance. We hypothesize that a small set of relevant
code observations and execution results can provide stronger diagnostic
support than a longer transcript containing repeated or untested repair
conjectures. This hypothesis could be tested by selectively retaining code
reads, reproduction outputs, test feedback, and recorded reasoning at
several fixed input budgets. Comparisons with equally long chronological
and randomly selected excerpts would separate evidence selection from
context length. Selection would use only information available before the
edit, while the checkpoint, Planner, and Executor remain fixed. Repeated
paired continuations would measure resolution, regressions, and planning
cost, testing whether selective evidence can preserve repair gains with
less computation.

\paragraph{Selective reuse of failed investigations.}
The recovery--cost trade-off between resuming and restarting raises a
further question about what should be retained from a failed attempt.
Experience-based agents extract reusable insights~\cite{Zhao2023ExpeLLA},
workflows~\cite{Wang2024AgentWM}, and grounded
plans~\cite{Hayashi2025SelfAbstractionFG} from past executions, offering
representations to compare when deciding what a repair attempt should retain.
We hypothesize that observed facts and reusable workspace artifacts provide
much of the reuse benefit, while earlier repair conjectures can sometimes
make it harder for the Executor to adopt a different plan. To examine this,
future experiments could compare the full history with a version that
masks prior diagnostic conjectures but retains executed repository
operations and their observations, keeping the workspace, supplied plan, and fresh
execution budget identical. A separate restart condition would reset both
history and workspace under the same plan and continuation budget.
Measuring repair outcomes, adoption of the new plan, and repeated
investigation would distinguish the effects of retained conjectures from
the broader cost of rebuilding state. Evaluating these hypotheses across
repository layouts and editing interfaces would test their generality.

\section{Conclusion}
\label{sec:conclusion}
We introduced \ourmodel{}, a framework that steers coding agents with a
single repair plan before their first source-code edit.
This action provides a search-free trigger for guidance grounded in the
agent's preceding investigation. A separate Planner supplies the plan,
and the same Executor continues with its accumulated context and workspace.
The framework supports both live steering and recovery from recorded failures.

Our evaluation on SWE-bench Verified and SWE-bench Multilingual shows that
one pre-edit plan can improve online resolution and recover failed attempts
offline. Online improvements across three harnesses and recovery with two
Executors support the applicability of this design, with gains varying by
Planner and execution setting. The results highlight complementary roles
for external planning and state reuse: suitable guidance can improve repair
outcomes, while retaining an existing investigation lowers the additional
Executor computation required for another attempt compared with restarting.
Comparisons with earlier and later pause positions find no point that
recovers more failures. The first source-code edit therefore offers a
practical opportunity to turn accumulated evidence into revised repair
decisions.

\section{Data Availability}
We will publicly release the implementation of \ourmodel{} for all three harnesses, the Planner prompt and configurations, per-task results and token counts for every table, the scripts that regenerate each table and figure, and the full agent trajectories.

\bibliographystyle{ACM-Reference-Format}
\bibliography{sample-base}


\begin{thebibliography}{50}


\ifx \showCODEN    \undefined \def \showCODEN     #1{\unskip}     \fi
\ifx \showDOI      \undefined \def \showDOI       #1{#1}\fi
\ifx \showISBNx    \undefined \def \showISBNx     #1{\unskip}     \fi
\ifx \showISBNxiii \undefined \def \showISBNxiii  #1{\unskip}     \fi
\ifx \showISSN     \undefined \def \showISSN      #1{\unskip}     \fi
\ifx \showLCCN     \undefined \def \showLCCN      #1{\unskip}     \fi
\ifx \shownote     \undefined \def \shownote      #1{#1}          \fi
\ifx \showarticletitle \undefined \def \showarticletitle #1{#1}   \fi
\ifx \showURL      \undefined \def \showURL       {\relax}        \fi
\providecommand\bibfield[2]{#2}
\providecommand\bibinfo[2]{#2}
\providecommand\natexlab[1]{#1}
\providecommand\showeprint[2][]{arXiv:#2}

\bibitem[Aggarwal et~al\mbox{.}(2025)]%
        {Aggarwal2025DARSDA}
\bibfield{author}{\bibinfo{person}{Vaibhav Aggarwal}, \bibinfo{person}{Ojasv
  Kamal}, \bibinfo{person}{Abhinav Japesh}, \bibinfo{person}{Zhijing Jin},
  {and} \bibinfo{person}{Bernhard Schölkopf}.}
  \bibinfo{year}{2025}\natexlab{}.
\newblock \showarticletitle{{DARS: Dynamic Action Re-Sampling to Enhance Coding
  Agent Performance by Adaptive Tree Traversal}}. In
  \bibinfo{booktitle}{\emph{Proceedings of the 63rd Annual Meeting of the
  Association for Computational Linguistics (Volume 1: Long Papers)}}.
  \bibinfo{pages}{19808--19855}.
\newblock
\urldef\tempurl%
\url{https://doi.org/10.18653/v1/2025.acl-long.973}
\showDOI{\tempurl}


\bibitem[Antoniades et~al\mbox{.}(2025)]%
        {antoniades2025swesearch}
\bibfield{author}{\bibinfo{person}{Antonis Antoniades}, \bibinfo{person}{Albert
  Örwall}, \bibinfo{person}{Kexun Zhang}, \bibinfo{person}{Yuxi Xie},
  \bibinfo{person}{Anirudh Goyal}, {and} \bibinfo{person}{William Wang}.}
  \bibinfo{year}{2025}\natexlab{}.
\newblock \showarticletitle{{SWE-Search: Enhancing Software Agents with Monte
  Carlo Tree Search and Iterative Refinement}}. In
  \bibinfo{booktitle}{\emph{International Conference on Learning
  Representations}}, Vol.~\bibinfo{volume}{2025}.
  \bibinfo{pages}{64485--64515}.
\newblock
\urldef\tempurl%
\url{https://proceedings.iclr.cc/paper_files/paper/2025/hash/a1e6783e4d739196cad3336f12d402bf-Abstract-Conference.html}
\showURL{%
\tempurl}


\bibitem[Bairi et~al\mbox{.}(2024)]%
        {Bairi2023CodePlanRC}
\bibfield{author}{\bibinfo{person}{Ramakrishna Bairi}, \bibinfo{person}{Atharv
  Sonwane}, \bibinfo{person}{Aditya Kanade}, \bibinfo{person}{Vageesh~D. C.},
  \bibinfo{person}{Arun Iyer}, \bibinfo{person}{Suresh Parthasarathy},
  \bibinfo{person}{Sriram Rajamani}, \bibinfo{person}{B. Ashok}, {and}
  \bibinfo{person}{Shashank Shet}.} \bibinfo{year}{2024}\natexlab{}.
\newblock \showarticletitle{{CodePlan: Repository-Level Coding using LLMs and
  Planning}}.
\newblock \bibinfo{journal}{\emph{Proceedings of the ACM on Software
  Engineering}} \bibinfo{volume}{1}, \bibinfo{number}{FSE}
  (\bibinfo{year}{2024}), \bibinfo{pages}{675--698}.
\newblock
\urldef\tempurl%
\url{https://doi.org/10.1145/3643757}
\showDOI{\tempurl}


\bibitem[Bouzenia et~al\mbox{.}(2025)]%
        {Bouzenia2024RepairAgentAA}
\bibfield{author}{\bibinfo{person}{Islem Bouzenia}, \bibinfo{person}{Premkumar
  Devanbu}, {and} \bibinfo{person}{Michael Pradel}.}
  \bibinfo{year}{2025}\natexlab{}.
\newblock \showarticletitle{{RepairAgent: An Autonomous, LLM-Based Agent for
  Program Repair}}. In \bibinfo{booktitle}{\emph{2025 IEEE/ACM 47th
  International Conference on Software Engineering (ICSE)}}.
  \bibinfo{pages}{2188--2200}.
\newblock
\urldef\tempurl%
\url{https://doi.org/10.1109/icse55347.2025.00157}
\showDOI{\tempurl}


\bibitem[Chen et~al\mbox{.}(2023)]%
        {Chen2022CodeTCG}
\bibfield{author}{\bibinfo{person}{Bei Chen}, \bibinfo{person}{Fengji Zhang},
  \bibinfo{person}{Anh Nguyen}, \bibinfo{person}{Daoguang Zan},
  \bibinfo{person}{Zeqi Lin}, \bibinfo{person}{Jian-Guang Lou}, {and}
  \bibinfo{person}{Weizhu Chen}.} \bibinfo{year}{2023}\natexlab{}.
\newblock \showarticletitle{{CodeT: Code Generation with Generated Tests}}. In
  \bibinfo{booktitle}{\emph{International Conference on Learning
  Representations}}.
\newblock
\urldef\tempurl%
\url{https://openreview.net/forum?id=ktrw68Cmu9c}
\showURL{%
\tempurl}


\bibitem[Chen et~al\mbox{.}(2024)]%
        {chen2023teaching}
\bibfield{author}{\bibinfo{person}{Xinyun Chen}, \bibinfo{person}{Maxwell Lin},
  \bibinfo{person}{Nathanael Schaerli}, {and} \bibinfo{person}{Denny Zhou}.}
  \bibinfo{year}{2024}\natexlab{}.
\newblock \showarticletitle{{Teaching Large Language Models to Self-Debug}}. In
  \bibinfo{booktitle}{\emph{International Conference on Learning
  Representations}}, Vol.~\bibinfo{volume}{2024}. \bibinfo{pages}{8746--8825}.
\newblock
\urldef\tempurl%
\url{https://proceedings.iclr.cc/paper_files/paper/2024/hash/2460396f2d0d421885997dd1612ac56b-Abstract-Conference.html}
\showURL{%
\tempurl}


\bibitem[Ding and Zhang(2026)]%
        {ding2026swereplay}
\bibfield{author}{\bibinfo{person}{Yifeng Ding} {and} \bibinfo{person}{Lingming
  Zhang}.} \bibinfo{year}{2026}\natexlab{}.
\newblock \showarticletitle{SWE-Replay: Efficient Test-Time Scaling for
  Software Engineering Agents}.
\newblock \bibinfo{journal}{\emph{ArXiv}}  \bibinfo{volume}{abs/2601.22129}
  (\bibinfo{year}{2026}).
\newblock


\bibitem[Dong et~al\mbox{.}(2024)]%
        {Dong2023SelfCollaborationCG}
\bibfield{author}{\bibinfo{person}{Yihong Dong}, \bibinfo{person}{Xue Jiang},
  \bibinfo{person}{Zhi Jin}, {and} \bibinfo{person}{Ge Li}.}
  \bibinfo{year}{2024}\natexlab{}.
\newblock \showarticletitle{{Self-Collaboration Code Generation via ChatGPT}}.
\newblock \bibinfo{journal}{\emph{ACM Transactions on Software Engineering and
  Methodology}} \bibinfo{volume}{33}, \bibinfo{number}{7}
  (\bibinfo{year}{2024}), \bibinfo{pages}{1--38}.
\newblock
\urldef\tempurl%
\url{https://doi.org/10.1145/3672459}
\showDOI{\tempurl}


\bibitem[Erdogan et~al\mbox{.}(2025)]%
        {Erdogan2025PlanandActIP}
\bibfield{author}{\bibinfo{person}{Lutfi~Eren Erdogan},
  \bibinfo{person}{Nicholas Lee}, \bibinfo{person}{Sehoon Kim},
  \bibinfo{person}{Suhong Moon}, \bibinfo{person}{Hiroki Furuta},
  \bibinfo{person}{Gopala Anumanchipalli}, \bibinfo{person}{Kurt Keutzer},
  {and} \bibinfo{person}{Amir Gholami}.} \bibinfo{year}{2025}\natexlab{}.
\newblock \showarticletitle{{Plan-and-Act: Improving Planning of Agents for
  Long-Horizon Tasks}}. In \bibinfo{booktitle}{\emph{Proceedings of the 42nd
  International Conference on Machine Learning}}, Vol.~\bibinfo{volume}{267}.
  \bibinfo{pages}{15419--15462}.
\newblock
\urldef\tempurl%
\url{https://proceedings.mlr.press/v267/erdogan25a.html}
\showURL{%
\tempurl}


\bibitem[Gandhi et~al\mbox{.}(2025a)]%
        {gandhi2025empirical}
\bibfield{author}{\bibinfo{person}{Shubham Gandhi}, \bibinfo{person}{Atharva
  Naik}, \bibinfo{person}{Yiqing Xie}, {and} \bibinfo{person}{Carolyn Rose}.}
  \bibinfo{year}{2025}\natexlab{a}.
\newblock \showarticletitle{{An Empirical Study on Strong-Weak Model
  Collaboration for Repo-level Code Generation}}. In
  \bibinfo{booktitle}{\emph{Proceedings of the 2025 Conference on Empirical
  Methods in Natural Language Processing}}. \bibinfo{pages}{20667--20686}.
\newblock
\urldef\tempurl%
\url{https://doi.org/10.18653/v1/2025.emnlp-main.1043}
\showDOI{\tempurl}


\bibitem[Gandhi et~al\mbox{.}(2025b)]%
        {gandhi2025swe}
\bibfield{author}{\bibinfo{person}{Shubham Gandhi}, \bibinfo{person}{Jason
  Tsay}, \bibinfo{person}{Jatin Ganhotra}, \bibinfo{person}{K. Kate}, {and}
  \bibinfo{person}{Yara Rizk}.} \bibinfo{year}{2025}\natexlab{b}.
\newblock \showarticletitle{When Agents go Astray: Course-Correcting SWE Agents
  with PRMs}.
\newblock \bibinfo{journal}{\emph{ArXiv}}  \bibinfo{volume}{abs/2509.02360}
  (\bibinfo{year}{2025}).
\newblock


\bibitem[Gou et~al\mbox{.}(2024)]%
        {gou2024critic}
\bibfield{author}{\bibinfo{person}{Zhibin Gou}, \bibinfo{person}{Zhihong Shao},
  \bibinfo{person}{Yeyun Gong}, \bibinfo{person}{yelong shen},
  \bibinfo{person}{Yujiu Yang}, \bibinfo{person}{Nan Duan}, {and}
  \bibinfo{person}{Weizhu Chen}.} \bibinfo{year}{2024}\natexlab{}.
\newblock \showarticletitle{{CRITIC: Large Language Models Can Self-Correct
  with Tool-Interactive Critiquing}}. In
  \bibinfo{booktitle}{\emph{International Conference on Learning
  Representations}}, Vol.~\bibinfo{volume}{2024}.
  \bibinfo{pages}{57734--57811}.
\newblock
\urldef\tempurl%
\url{https://proceedings.iclr.cc/paper_files/paper/2024/hash/fef126561bbf9d4467dbb8d27334b8fe-Abstract-Conference.html}
\showURL{%
\tempurl}


\bibitem[Guo et~al\mbox{.}(2025)]%
        {Lin2025SEAgentST}
\bibfield{author}{\bibinfo{person}{Yifu Guo}, \bibinfo{person}{Jiaye Lin},
  \bibinfo{person}{Huacan Wang}, \bibinfo{person}{Yuzhen Han},
  \bibinfo{person}{Sen Hu}, \bibinfo{person}{Ziyi Ni}, \bibinfo{person}{Licheng
  Wang}, {and} \bibinfo{person}{Mingguang Chen}.}
  \bibinfo{year}{2025}\natexlab{}.
\newblock \showarticletitle{{SE-Agent: Self-Evolution Trajectory Optimization
  in Multi-Step Reasoning with LLM-Based Agents}}. In
  \bibinfo{booktitle}{\emph{Advances in Neural Information Processing
  Systems}}, Vol.~\bibinfo{volume}{38, Main Conference}.
  \bibinfo{pages}{116314--116341}.
\newblock
\urldef\tempurl%
\url{https://doi.org/10.52202/085713-3877}
\showDOI{\tempurl}


\bibitem[Hayashi et~al\mbox{.}(2025)]%
        {Hayashi2025SelfAbstractionFG}
\bibfield{author}{\bibinfo{person}{Hiroaki Hayashi}, \bibinfo{person}{Bo Pang},
  \bibinfo{person}{Wenting Zhao}, \bibinfo{person}{Ye Liu},
  \bibinfo{person}{Akash Gokul}, \bibinfo{person}{Srijan Bansal},
  \bibinfo{person}{Cai-Ming Xiong}, \bibinfo{person}{Semih Yavuz}, {and}
  \bibinfo{person}{Ying-Bo Zhou}.} \bibinfo{year}{2025}\natexlab{}.
\newblock \showarticletitle{Self-Abstraction from Grounded Experience for
  Plan-Guided Policy Refinement}.
\newblock \bibinfo{journal}{\emph{ArXiv}}  \bibinfo{volume}{abs/2511.05931}
  (\bibinfo{year}{2025}).
\newblock


\bibitem[Hong et~al\mbox{.}(2024)]%
        {Hong2023MetaGPTMP}
\bibfield{author}{\bibinfo{person}{Sirui Hong}, \bibinfo{person}{Mingchen
  Zhuge}, \bibinfo{person}{Jonathan Chen}, \bibinfo{person}{Xiawu Zheng},
  \bibinfo{person}{Yuheng Cheng}, \bibinfo{person}{Jinlin Wang},
  \bibinfo{person}{Ceyao Zhang}, \bibinfo{person}{zili wang},
  \bibinfo{person}{Steven Yau}, \bibinfo{person}{Zijuan Lin},
  \bibinfo{person}{Liyang Zhou}, \bibinfo{person}{Chenyu Ran},
  \bibinfo{person}{Lingfeng Xiao}, \bibinfo{person}{Chenglin Wu}, {and}
  \bibinfo{person}{J\"{u}rgen Schmidhuber}.} \bibinfo{year}{2024}\natexlab{}.
\newblock \showarticletitle{{MetaGPT: Meta Programming for A Multi-Agent
  Collaborative Framework}}. In \bibinfo{booktitle}{\emph{International
  Conference on Learning Representations}}, Vol.~\bibinfo{volume}{2024}.
  \bibinfo{pages}{23247--23275}.
\newblock
\urldef\tempurl%
\url{https://proceedings.iclr.cc/paper_files/paper/2024/hash/6507b115562bb0a305f1958ccc87355a-Abstract-Conference.html}
\showURL{%
\tempurl}


\bibitem[Huang et~al\mbox{.}(2024)]%
        {huang2024large}
\bibfield{author}{\bibinfo{person}{Jie Huang}, \bibinfo{person}{Xinyun Chen},
  \bibinfo{person}{Swaroop Mishra}, \bibinfo{person}{Huaixiu~Steven Zheng},
  \bibinfo{person}{Adams Yu}, \bibinfo{person}{Xinying Song}, {and}
  \bibinfo{person}{Denny Zhou}.} \bibinfo{year}{2024}\natexlab{}.
\newblock \showarticletitle{{Large Language Models Cannot Self-Correct
  Reasoning Yet}}. In \bibinfo{booktitle}{\emph{International Conference on
  Learning Representations}}, Vol.~\bibinfo{volume}{2024}.
  \bibinfo{pages}{32808--32824}.
\newblock
\urldef\tempurl%
\url{https://proceedings.iclr.cc/paper_files/paper/2024/hash/8b4add8b0aa8749d80a34ca5d941c355-Abstract-Conference.html}
\showURL{%
\tempurl}


\bibitem[Jimenez et~al\mbox{.}(2024)]%
        {jimenez2024swebench}
\bibfield{author}{\bibinfo{person}{Carlos~E Jimenez}, \bibinfo{person}{John
  Yang}, \bibinfo{person}{Alexander Wettig}, \bibinfo{person}{Shunyu Yao},
  \bibinfo{person}{Kexin Pei}, \bibinfo{person}{Ofir Press}, {and}
  \bibinfo{person}{Karthik Narasimhan}.} \bibinfo{year}{2024}\natexlab{}.
\newblock \showarticletitle{{SWE-bench: Can Language Models Resolve Real-world
  Github Issues?}}. In \bibinfo{booktitle}{\emph{International Conference on
  Learning Representations}}, Vol.~\bibinfo{volume}{2024}.
  \bibinfo{pages}{54107--54157}.
\newblock
\urldef\tempurl%
\url{https://proceedings.iclr.cc/paper_files/paper/2024/hash/edac78c3e300629acfe6cbe9ca88fb84-Abstract-Conference.html}
\showURL{%
\tempurl}


\bibitem[Kamoi et~al\mbox{.}(2024)]%
        {kamoi2024can}
\bibfield{author}{\bibinfo{person}{Ryo Kamoi}, \bibinfo{person}{Yusen Zhang},
  \bibinfo{person}{Nan Zhang}, \bibinfo{person}{Jiawei Han}, {and}
  \bibinfo{person}{Rui Zhang}.} \bibinfo{year}{2024}\natexlab{}.
\newblock \showarticletitle{{When Can {LLM}s Actually Correct Their Own
  Mistakes? A Critical Survey of Self-Correction of {LLM}s}}.
\newblock \bibinfo{journal}{\emph{Transactions of the Association for
  Computational Linguistics}}  \bibinfo{volume}{12} (\bibinfo{year}{2024}),
  \bibinfo{pages}{1417--1440}.
\newblock
\urldef\tempurl%
\url{https://doi.org/10.1162/tacl\_a\_00713}
\showDOI{\tempurl}


\bibitem[Lee et~al\mbox{.}(2026)]%
        {lee2026cope}
\bibfield{author}{\bibinfo{person}{Byeongchan Lee}, \bibinfo{person}{Jonghoon
  Lee}, \bibinfo{person}{Dongyoung Kim}, \bibinfo{person}{Jaehyung Kim},
  \bibinfo{person}{Kyungjoon Park}, \bibinfo{person}{Dongjun Lee}, {and}
  \bibinfo{person}{Jinwoo Shin}.} \bibinfo{year}{2026}\natexlab{}.
\newblock \showarticletitle{{Efficient {LLM} Collaboration via Planning}}.
\newblock \bibinfo{journal}{\emph{Transactions on Machine Learning Research}}
  (\bibinfo{year}{2026}).
\newblock
\urldef\tempurl%
\url{https://openreview.net/forum?id=RPzbeL0koP}
\showURL{%
\tempurl}


\bibitem[Li et~al\mbox{.}(2025)]%
        {Li2024CodeTreeAT}
\bibfield{author}{\bibinfo{person}{Jierui Li}, \bibinfo{person}{Hung Le},
  \bibinfo{person}{Yingbo Zhou}, \bibinfo{person}{Caiming Xiong},
  \bibinfo{person}{Silvio Savarese}, {and} \bibinfo{person}{Doyen Sahoo}.}
  \bibinfo{year}{2025}\natexlab{}.
\newblock \showarticletitle{{CodeTree: Agent-guided Tree Search for Code
  Generation with Large Language Models}}. In
  \bibinfo{booktitle}{\emph{Proceedings of the 2025 Conference of the Nations
  of the Americas Chapter of the Association for Computational Linguistics:
  Human Language Technologies (Volume 1: Long Papers)}}.
  \bibinfo{pages}{3711--3726}.
\newblock
\urldef\tempurl%
\url{https://doi.org/10.18653/v1/2025.naacl-long.189}
\showDOI{\tempurl}


\bibitem[Liu et~al\mbox{.}(2026a)]%
        {Liu2025ProcessCentricAO}
\bibfield{author}{\bibinfo{person}{Shuyang Liu}, \bibinfo{person}{Yang Chen},
  \bibinfo{person}{Rahul Krishna}, \bibinfo{person}{Saurabh Sinha},
  \bibinfo{person}{Jatin Ganhotra}, {and} \bibinfo{person}{Reyhaneh
  Jabbarvand}.} \bibinfo{year}{2026}\natexlab{a}.
\newblock \showarticletitle{{Process-Centric Analysis of Agentic Software
  Systems}}.
\newblock \bibinfo{journal}{\emph{Proceedings of the ACM on Programming
  Languages}} \bibinfo{volume}{10}, \bibinfo{number}{OOPSLA1}
  (\bibinfo{year}{2026}), \bibinfo{pages}{1961--1988}.
\newblock
\urldef\tempurl%
\url{https://doi.org/10.1145/3798271}
\showDOI{\tempurl}


\bibitem[Liu et~al\mbox{.}(2026b)]%
        {liu2026liveplan}
\bibfield{author}{\bibinfo{person}{Shuyang Liu}, \bibinfo{person}{Saman
  Dehghan}, \bibinfo{person}{Ji~Young Kim}, \bibinfo{person}{Jatin Ganhotra},
  \bibinfo{person}{Martin Hirzel}, {and} \bibinfo{person}{Reyhaneh
  Jabbarvand}.} \bibinfo{year}{2026}\natexlab{b}.
\newblock \bibinfo{title}{Online Monitoring and Corrective Steering of
  Programming Agents}.
\newblock
\newblock
\showeprint[arxiv]{2608.06701}
\urldef\tempurl%
\url{https://arxiv.org/abs/2608.06701}
\showURL{%
\tempurl}


\bibitem[Ma et~al\mbox{.}(2025)]%
        {Ma2024AlibabaLI}
\bibfield{author}{\bibinfo{person}{Yingwei Ma}, \bibinfo{person}{Qingping
  Yang}, \bibinfo{person}{Rongyu Cao}, \bibinfo{person}{Binhua Li},
  \bibinfo{person}{Fei Huang}, {and} \bibinfo{person}{Yongbin Li}.}
  \bibinfo{year}{2025}\natexlab{}.
\newblock \showarticletitle{{Alibaba LingmaAgent: Improving Automated Issue
  Resolution via Comprehensive Repository Exploration}}. In
  \bibinfo{booktitle}{\emph{Proceedings of the 33rd ACM International
  Conference on the Foundations of Software Engineering}}.
  \bibinfo{pages}{238--249}.
\newblock
\urldef\tempurl%
\url{https://doi.org/10.1145/3696630.3728549}
\showDOI{\tempurl}


\bibitem[Madaan et~al\mbox{.}(2023)]%
        {madaan2023selfrefine}
\bibfield{author}{\bibinfo{person}{Aman Madaan}, \bibinfo{person}{Niket
  Tandon}, \bibinfo{person}{Prakhar Gupta}, \bibinfo{person}{Skyler Hallinan},
  \bibinfo{person}{Luyu Gao}, \bibinfo{person}{Sarah Wiegreffe},
  \bibinfo{person}{Uri Alon}, \bibinfo{person}{Nouha Dziri},
  \bibinfo{person}{Shrimai Prabhumoye}, \bibinfo{person}{Yiming Yang},
  \bibinfo{person}{Shashank Gupta}, \bibinfo{person}{Bodhisattwa~Prasad
  Majumder}, \bibinfo{person}{Katherine Hermann}, \bibinfo{person}{Sean
  Welleck}, \bibinfo{person}{Amir Yazdanbakhsh}, {and} \bibinfo{person}{Peter
  Clark}.} \bibinfo{year}{2023}\natexlab{}.
\newblock \showarticletitle{{Self-Refine: Iterative Refinement with
  Self-Feedback}}. In \bibinfo{booktitle}{\emph{Advances in Neural Information
  Processing Systems}}, Vol.~\bibinfo{volume}{36}.
  \bibinfo{pages}{46534--46594}.
\newblock
\urldef\tempurl%
\url{https://doi.org/10.52202/075280-2019}
\showDOI{\tempurl}


\bibitem[Nanda et~al\mbox{.}(2026)]%
        {Nanda2026WinkRF}
\bibfield{author}{\bibinfo{person}{Rahul Nanda}, \bibinfo{person}{Chandra
  Maddila}, \bibinfo{person}{Smriti Jha}, \bibinfo{person}{Euna~Mehnaz Khan},
  \bibinfo{person}{Matteo Paltenghi}, {and} \bibinfo{person}{Satish Chandra}.}
  \bibinfo{year}{2026}\natexlab{}.
\newblock \showarticletitle{{Wink: Recovering from Misbehaviors in Coding
  Agents}}. In \bibinfo{booktitle}{\emph{Proceedings of the 3rd ACM
  International Conference on AI-Powered Software}}. \bibinfo{pages}{208--217}.
\newblock
\urldef\tempurl%
\url{https://doi.org/10.1145/3805760.3814911}
\showDOI{\tempurl}


\bibitem[Ni et~al\mbox{.}(2023)]%
        {Ni2023LEVERLT}
\bibfield{author}{\bibinfo{person}{Ansong Ni}, \bibinfo{person}{Srini Iyer},
  \bibinfo{person}{Dragomir Radev}, \bibinfo{person}{Veselin Stoyanov},
  \bibinfo{person}{Wen-Tau Yih}, \bibinfo{person}{Sida Wang}, {and}
  \bibinfo{person}{Xi~Victoria Lin}.} \bibinfo{year}{2023}\natexlab{}.
\newblock \showarticletitle{{{LEVER}: Learning to Verify Language-to-Code
  Generation with Execution}}. In \bibinfo{booktitle}{\emph{Proceedings of the
  40th International Conference on Machine Learning}},
  Vol.~\bibinfo{volume}{202}. \bibinfo{pages}{26106--26128}.
\newblock
\urldef\tempurl%
\url{https://proceedings.mlr.press/v202/ni23b.html}
\showURL{%
\tempurl}


\bibitem[Ouyang et~al\mbox{.}(2025)]%
        {Ouyang2024RepoGraphEA}
\bibfield{author}{\bibinfo{person}{Siru Ouyang}, \bibinfo{person}{Wenhao Yu},
  \bibinfo{person}{Kaixin Ma}, \bibinfo{person}{Zilin Xiao},
  \bibinfo{person}{Zhihan Zhang}, \bibinfo{person}{Mengzhao Jia},
  \bibinfo{person}{Jiawei Han}, \bibinfo{person}{Hongming Zhang}, {and}
  \bibinfo{person}{Dong Yu}.} \bibinfo{year}{2025}\natexlab{}.
\newblock \showarticletitle{{RepoGraph: Enhancing AI Software Engineering with
  Repository-level Code Graph}}. In \bibinfo{booktitle}{\emph{International
  Conference on Learning Representations}}, Vol.~\bibinfo{volume}{2025}.
  \bibinfo{pages}{30098--30121}.
\newblock
\urldef\tempurl%
\url{https://proceedings.iclr.cc/paper_files/paper/2025/hash/4a4a3c197deac042461c677219efd36c-Abstract-Conference.html}
\showURL{%
\tempurl}


\bibitem[Pan et~al\mbox{.}(2025)]%
        {Pan2024TrainingSE}
\bibfield{author}{\bibinfo{person}{Jiayi Pan}, \bibinfo{person}{Xingyao Wang},
  \bibinfo{person}{Graham Neubig}, \bibinfo{person}{Navdeep Jaitly},
  \bibinfo{person}{Heng Ji}, \bibinfo{person}{Alane Suhr}, {and}
  \bibinfo{person}{Yizhe Zhang}.} \bibinfo{year}{2025}\natexlab{}.
\newblock \showarticletitle{{Training Software Engineering Agents and Verifiers
  with SWE-Gym}}. In \bibinfo{booktitle}{\emph{Proceedings of the 42nd
  International Conference on Machine Learning}}, Vol.~\bibinfo{volume}{267}.
  \bibinfo{pages}{47717--47737}.
\newblock
\urldef\tempurl%
\url{https://proceedings.mlr.press/v267/pan25g.html}
\showURL{%
\tempurl}


\bibitem[Robeyns et~al\mbox{.}(2025)]%
        {robeyns2025sica}
\bibfield{author}{\bibinfo{person}{M. Robeyns}, \bibinfo{person}{M. Szummer},
  {and} \bibinfo{person}{Laurence Aitchison}.} \bibinfo{year}{2025}\natexlab{}.
\newblock \showarticletitle{A Self-Improving Coding Agent}.
\newblock \bibinfo{journal}{\emph{ArXiv}}  \bibinfo{volume}{abs/2504.15228}
  (\bibinfo{year}{2025}).
\newblock


\bibitem[Ruan et~al\mbox{.}(2025)]%
        {Ruan2024SpecRoverCI}
\bibfield{author}{\bibinfo{person}{Haifeng Ruan}, \bibinfo{person}{Yuntong
  Zhang}, {and} \bibinfo{person}{Abhik Roychoudhury}.}
  \bibinfo{year}{2025}\natexlab{}.
\newblock \showarticletitle{{SpecRover: Code Intent Extraction via LLMs}}. In
  \bibinfo{booktitle}{\emph{2025 IEEE/ACM 47th International Conference on
  Software Engineering (ICSE)}}. \bibinfo{pages}{963--974}.
\newblock
\urldef\tempurl%
\url{https://doi.org/10.1109/icse55347.2025.00080}
\showDOI{\tempurl}


\bibitem[Shinn et~al\mbox{.}(2023)]%
        {shinn2023reflexion}
\bibfield{author}{\bibinfo{person}{Noah Shinn}, \bibinfo{person}{Federico
  Cassano}, \bibinfo{person}{Ashwin Gopinath}, \bibinfo{person}{Karthik
  Narasimhan}, {and} \bibinfo{person}{Shunyu Yao}.}
  \bibinfo{year}{2023}\natexlab{}.
\newblock \showarticletitle{{Reflexion: language agents with verbal
  reinforcement learning}}. In \bibinfo{booktitle}{\emph{Advances in Neural
  Information Processing Systems}}, Vol.~\bibinfo{volume}{36}.
  \bibinfo{pages}{8634--8652}.
\newblock
\urldef\tempurl%
\url{https://doi.org/10.52202/075280-0377}
\showDOI{\tempurl}


\bibitem[Tamoyan et~al\mbox{.}(2026)]%
        {tamoyan2026sherloc}
\bibfield{author}{\bibinfo{person}{Hovhannes Tamoyan}, \bibinfo{person}{Sean
  Narenthiran}, \bibinfo{person}{Erik Arakelyan}, \bibinfo{person}{Mira
  Mezini}, {and} \bibinfo{person}{Boris Ginsburg}.}
  \bibinfo{year}{2026}\natexlab{}.
\newblock \showarticletitle{SHERLOC: Structured Diagnostic Localization for
  Code Repair Agents}.
\newblock \bibinfo{journal}{\emph{ArXiv}}  \bibinfo{volume}{abs/2606.24820}
  (\bibinfo{year}{2026}).
\newblock


\bibitem[Tan et~al\mbox{.}(2026)]%
        {tan2026swemanager}
\bibfield{author}{\bibinfo{person}{Boyin Tan}, \bibinfo{person}{Haoning Deng},
  \bibinfo{person}{Junyuan Zhang}, \bibinfo{person}{Junjielong Xu},
  \bibinfo{person}{Pinjia He}, {and} \bibinfo{person}{Youcheng Sun}.}
  \bibinfo{year}{2026}\natexlab{}.
\newblock \showarticletitle{SWE-Manager: Selecting and Synthesizing Golden
  Proposals Before Coding}.
\newblock \bibinfo{journal}{\emph{ArXiv}}  \bibinfo{volume}{abs/2601.22956}
  (\bibinfo{year}{2026}).
\newblock


\bibitem[Tao et~al\mbox{.}(2024)]%
        {Tao2024MAGISLM}
\bibfield{author}{\bibinfo{person}{Wei Tao}, \bibinfo{person}{Yucheng Zhou},
  \bibinfo{person}{Yanlin Wang}, \bibinfo{person}{Wenqiang Zhang},
  \bibinfo{person}{Hongyu Zhang}, {and} \bibinfo{person}{Yu Cheng}.}
  \bibinfo{year}{2024}\natexlab{}.
\newblock \showarticletitle{{MAGIS: LLM-Based Multi-Agent Framework for GitHub
  Issue Resolution}}. In \bibinfo{booktitle}{\emph{Advances in Neural
  Information Processing Systems}}, Vol.~\bibinfo{volume}{37}.
  \bibinfo{pages}{51963--51993}.
\newblock
\urldef\tempurl%
\url{https://doi.org/10.52202/079017-1647}
\showDOI{\tempurl}


\bibitem[Wang et~al\mbox{.}(2023)]%
        {Wang2023PlanandSolvePI}
\bibfield{author}{\bibinfo{person}{Lei Wang}, \bibinfo{person}{Wanyu Xu},
  \bibinfo{person}{Yihuai Lan}, \bibinfo{person}{Zhiqiang Hu},
  \bibinfo{person}{Yunshi Lan}, \bibinfo{person}{Roy Ka-Wei Lee}, {and}
  \bibinfo{person}{Ee-Peng Lim}.} \bibinfo{year}{2023}\natexlab{}.
\newblock \showarticletitle{{Plan-and-Solve Prompting: Improving Zero-Shot
  Chain-of-Thought Reasoning by Large Language Models}}. In
  \bibinfo{booktitle}{\emph{Proceedings of the 61st Annual Meeting of the
  Association for Computational Linguistics (Volume 1: Long Papers)}}.
  \bibinfo{pages}{2609--2634}.
\newblock
\urldef\tempurl%
\url{https://doi.org/10.18653/v1/2023.acl-long.147}
\showDOI{\tempurl}


\bibitem[Wang et~al\mbox{.}(2025a)]%
        {wang2024openhands}
\bibfield{author}{\bibinfo{person}{Xingyao Wang}, \bibinfo{person}{Boxuan Li},
  \bibinfo{person}{Yufan Song}, \bibinfo{person}{Frank~F Xu},
  \bibinfo{person}{Xiangru Tang}, \bibinfo{person}{Mingchen Zhuge},
  \bibinfo{person}{Jiayi Pan}, \bibinfo{person}{Yueqi Song},
  \bibinfo{person}{Bowen Li}, \bibinfo{person}{Jaskirat Singh},
  \bibinfo{person}{Hoang Tran}, \bibinfo{person}{Fuqiang Li},
  \bibinfo{person}{Ren Ma}, \bibinfo{person}{Mingzhang Zheng},
  \bibinfo{person}{Bill Qian}, \bibinfo{person}{Daniel Shao},
  \bibinfo{person}{Niklas Muennighoff}, \bibinfo{person}{Yizhe Zhang},
  \bibinfo{person}{Binyuan Hui}, \bibinfo{person}{Junyang Lin},
  \bibinfo{person}{Robert Brennan}, \bibinfo{person}{Hao Peng},
  \bibinfo{person}{Heng Ji}, {and} \bibinfo{person}{Graham Neubig}.}
  \bibinfo{year}{2025}\natexlab{a}.
\newblock \showarticletitle{{OpenHands: An Open Platform for AI Software
  Developers as Generalist Agents}}. In \bibinfo{booktitle}{\emph{International
  Conference on Learning Representations}}, Vol.~\bibinfo{volume}{2025}.
  \bibinfo{pages}{65882--65919}.
\newblock
\urldef\tempurl%
\url{https://proceedings.iclr.cc/paper_files/paper/2025/hash/a4b6ad6b48850c0c331d1259fc66a69c-Abstract-Conference.html}
\showURL{%
\tempurl}


\bibitem[Wang et~al\mbox{.}(2025b)]%
        {Wang2024AgentWM}
\bibfield{author}{\bibinfo{person}{Zora~Zhiruo Wang}, \bibinfo{person}{Jiayuan
  Mao}, \bibinfo{person}{Daniel Fried}, {and} \bibinfo{person}{Graham Neubig}.}
  \bibinfo{year}{2025}\natexlab{b}.
\newblock \showarticletitle{{Agent Workflow Memory}}. In
  \bibinfo{booktitle}{\emph{Proceedings of the 42nd International Conference on
  Machine Learning}}, Vol.~\bibinfo{volume}{267}.
  \bibinfo{pages}{63897--63911}.
\newblock
\urldef\tempurl%
\url{https://proceedings.mlr.press/v267/wang25bx.html}
\showURL{%
\tempurl}


\bibitem[Xia et~al\mbox{.}(2025)]%
        {xia2024agentless}
\bibfield{author}{\bibinfo{person}{Chunqiu~Steven Xia}, \bibinfo{person}{Yinlin
  Deng}, \bibinfo{person}{Soren Dunn}, {and} \bibinfo{person}{Lingming Zhang}.}
  \bibinfo{year}{2025}\natexlab{}.
\newblock \showarticletitle{{Demystifying LLM-Based Software Engineering
  Agents}}.
\newblock \bibinfo{journal}{\emph{Proceedings of the ACM on Software
  Engineering}} \bibinfo{volume}{2}, \bibinfo{number}{FSE}
  (\bibinfo{year}{2025}), \bibinfo{pages}{801--824}.
\newblock
\urldef\tempurl%
\url{https://doi.org/10.1145/3715754}
\showDOI{\tempurl}


\bibitem[Xia and Zhang(2024)]%
        {Xia2023AutomatedPR}
\bibfield{author}{\bibinfo{person}{Chunqiu~Steven Xia} {and}
  \bibinfo{person}{Lingming Zhang}.} \bibinfo{year}{2024}\natexlab{}.
\newblock \showarticletitle{{Automated Program Repair via Conversation: Fixing
  162 out of 337 Bugs for \$0.42 Each using ChatGPT}}. In
  \bibinfo{booktitle}{\emph{ISSTA 2024 - Proceedings of the 33rd ACM SIGSOFT
  International Symposium on Software Testing and Analysis}}.
  \bibinfo{pages}{819--831}.
\newblock
\urldef\tempurl%
\url{https://doi.org/10.1145/3650212.3680323}
\showDOI{\tempurl}


\bibitem[Xie et~al\mbox{.}(2025)]%
        {Xie2025SWEFixerTO}
\bibfield{author}{\bibinfo{person}{Chengxing Xie}, \bibinfo{person}{Bowen Li},
  \bibinfo{person}{Chang Gao}, \bibinfo{person}{He Du}, \bibinfo{person}{Wai
  Lam}, \bibinfo{person}{Difan Zou}, {and} \bibinfo{person}{Kai Chen}.}
  \bibinfo{year}{2025}\natexlab{}.
\newblock \showarticletitle{{SWE-Fixer: Training Open-Source LLMs for Effective
  and Efficient GitHub Issue Resolution}}. In
  \bibinfo{booktitle}{\emph{Findings of the Association for Computational
  Linguistics: ACL 2025}}. \bibinfo{pages}{1123--1139}.
\newblock
\urldef\tempurl%
\url{https://doi.org/10.18653/v1/2025.findings-acl.62}
\showDOI{\tempurl}


\bibitem[Yang et~al\mbox{.}(2024)]%
        {yang2024sweagent}
\bibfield{author}{\bibinfo{person}{John Yang}, \bibinfo{person}{Carlos
  Jimenez}, \bibinfo{person}{Alexander Wettig}, \bibinfo{person}{Kilian
  Lieret}, \bibinfo{person}{Shunyu Yao}, \bibinfo{person}{Karthik Narasimhan},
  {and} \bibinfo{person}{Ofir Press}.} \bibinfo{year}{2024}\natexlab{}.
\newblock \showarticletitle{{SWE-agent: Agent-Computer Interfaces Enable
  Automated Software Engineering}}. In \bibinfo{booktitle}{\emph{Advances in
  Neural Information Processing Systems}}, Vol.~\bibinfo{volume}{37}.
  \bibinfo{pages}{50528--50652}.
\newblock
\urldef\tempurl%
\url{https://doi.org/10.52202/079017-1601}
\showDOI{\tempurl}


\bibitem[Yang et~al\mbox{.}(2025)]%
        {yang2025swesmith}
\bibfield{author}{\bibinfo{person}{John Yang}, \bibinfo{person}{Kilian Lieret},
  \bibinfo{person}{Carlos Jimenez}, \bibinfo{person}{Alexander Wettig},
  \bibinfo{person}{Kabir Khandpur}, \bibinfo{person}{Yanzhe Zhang},
  \bibinfo{person}{Binyuan Hui}, \bibinfo{person}{Ofir Press},
  \bibinfo{person}{Ludwig Schmidt}, {and} \bibinfo{person}{Diyi Yang}.}
  \bibinfo{year}{2025}\natexlab{}.
\newblock \showarticletitle{{SWE-smith: Scaling Data for Software Engineering
  Agents}}. In \bibinfo{booktitle}{\emph{Advances in Neural Information
  Processing Systems}}, Vol.~\bibinfo{volume}{38, Main Conference}.
\newblock
\urldef\tempurl%
\url{https://doi.org/10.52202/085713-3239}
\showDOI{\tempurl}


\bibitem[Yao et~al\mbox{.}(2023a)]%
        {Yao2023TreeOT}
\bibfield{author}{\bibinfo{person}{Shunyu Yao}, \bibinfo{person}{Dian Yu},
  \bibinfo{person}{Jeffrey Zhao}, \bibinfo{person}{Izhak Shafran},
  \bibinfo{person}{Tom Griffiths}, \bibinfo{person}{Yuan Cao}, {and}
  \bibinfo{person}{Karthik Narasimhan}.} \bibinfo{year}{2023}\natexlab{a}.
\newblock \showarticletitle{{Tree of Thoughts: Deliberate Problem Solving with
  Large Language Models}}. In \bibinfo{booktitle}{\emph{Advances in Neural
  Information Processing Systems}}, Vol.~\bibinfo{volume}{36}.
  \bibinfo{pages}{11809--11822}.
\newblock
\urldef\tempurl%
\url{https://doi.org/10.52202/075280-0517}
\showDOI{\tempurl}


\bibitem[Yao et~al\mbox{.}(2023b)]%
        {yao2023react}
\bibfield{author}{\bibinfo{person}{Shunyu Yao}, \bibinfo{person}{Jeffrey Zhao},
  \bibinfo{person}{Dian Yu}, \bibinfo{person}{Nan Du}, \bibinfo{person}{Izhak
  Shafran}, \bibinfo{person}{Karthik Narasimhan}, {and} \bibinfo{person}{Yuan
  Cao}.} \bibinfo{year}{2023}\natexlab{b}.
\newblock \showarticletitle{{ReAct: Synergizing Reasoning and Acting in
  Language Models}}. In \bibinfo{booktitle}{\emph{International Conference on
  Learning Representations}}.
\newblock
\urldef\tempurl%
\url{https://openreview.net/forum?id=WE_vluYUL-X}
\showURL{%
\tempurl}


\bibitem[Zechner(2025)]%
        {zechner2025pi}
\bibfield{author}{\bibinfo{person}{Mario Zechner}.}
  \bibinfo{year}{2025}\natexlab{}.
\newblock \bibinfo{title}{{Pi}: A Minimal Terminal Coding Agent}.
\newblock
\newblock
\urldef\tempurl%
\url{https://github.com/badlogic/pi-mono}
\showURL{%
\tempurl}


\bibitem[Zhang et~al\mbox{.}(2024a)]%
        {Zhang2024CodeAgentEC}
\bibfield{author}{\bibinfo{person}{Kechi Zhang}, \bibinfo{person}{Jia Li},
  \bibinfo{person}{Ge Li}, \bibinfo{person}{Xianjie Shi}, {and}
  \bibinfo{person}{Zhi Jin}.} \bibinfo{year}{2024}\natexlab{a}.
\newblock \showarticletitle{{{C}ode{A}gent: Enhancing Code Generation with
  Tool-Integrated Agent Systems for Real-World Repo-level Coding Challenges}}.
  In \bibinfo{booktitle}{\emph{Proceedings of the 62nd Annual Meeting of the
  Association for Computational Linguistics (Volume 1: Long Papers)}}.
  \bibinfo{pages}{13643--13658}.
\newblock
\urldef\tempurl%
\url{https://doi.org/10.18653/v1/2024.acl-long.737}
\showDOI{\tempurl}


\bibitem[Zhang et~al\mbox{.}(2024b)]%
        {zhang2024autocoderover}
\bibfield{author}{\bibinfo{person}{Yuntong Zhang}, \bibinfo{person}{Haifeng
  Ruan}, \bibinfo{person}{Zhiyu Fan}, {and} \bibinfo{person}{Abhik
  Roychoudhury}.} \bibinfo{year}{2024}\natexlab{b}.
\newblock \showarticletitle{{AutoCodeRover: Autonomous Program Improvement}}.
  In \bibinfo{booktitle}{\emph{Proceedings of the 33rd ACM SIGSOFT
  International Symposium on Software Testing and Analysis}}.
  \bibinfo{pages}{1592--1604}.
\newblock
\urldef\tempurl%
\url{https://doi.org/10.1145/3650212.3680384}
\showDOI{\tempurl}


\bibitem[Zhao et~al\mbox{.}(2024)]%
        {Zhao2023ExpeLLA}
\bibfield{author}{\bibinfo{person}{Andrew Zhao}, \bibinfo{person}{Daniel
  Huang}, \bibinfo{person}{Quentin Xu}, \bibinfo{person}{Matthieu Lin},
  \bibinfo{person}{Yong-Jin Liu}, {and} \bibinfo{person}{Gao Huang}.}
  \bibinfo{year}{2024}\natexlab{}.
\newblock \showarticletitle{{ExpeL: LLM Agents Are Experiential Learners}}.
\newblock \bibinfo{journal}{\emph{Proceedings of the AAAI Conference on
  Artificial Intelligence}} \bibinfo{volume}{38}, \bibinfo{number}{17}
  (\bibinfo{year}{2024}), \bibinfo{pages}{19632--19642}.
\newblock
\urldef\tempurl%
\url{https://doi.org/10.1609/aaai.v38i17.29936}
\showDOI{\tempurl}


\bibitem[Zhong et~al\mbox{.}(2024)]%
        {Zhong2024DebugLA}
\bibfield{author}{\bibinfo{person}{Li Zhong}, \bibinfo{person}{Zilong Wang},
  {and} \bibinfo{person}{Jingbo Shang}.} \bibinfo{year}{2024}\natexlab{}.
\newblock \showarticletitle{{Debug like a Human: A Large Language Model
  Debugger via Verifying Runtime Execution Step by Step}}. In
  \bibinfo{booktitle}{\emph{Findings of the Association for Computational
  Linguistics: ACL 2024}}. \bibinfo{pages}{851--870}.
\newblock
\urldef\tempurl%
\url{https://doi.org/10.18653/v1/2024.findings-acl.49}
\showDOI{\tempurl}


\bibitem[Zhou et~al\mbox{.}(2024)]%
        {Zhou2023LanguageAT}
\bibfield{author}{\bibinfo{person}{Andy Zhou}, \bibinfo{person}{Kai Yan},
  \bibinfo{person}{Michal Shlapentokh-Rothman}, \bibinfo{person}{Haohan Wang},
  {and} \bibinfo{person}{Yu-Xiong Wang}.} \bibinfo{year}{2024}\natexlab{}.
\newblock \showarticletitle{{Language Agent Tree Search Unifies Reasoning,
  Acting, and Planning in Language Models}}. In
  \bibinfo{booktitle}{\emph{Proceedings of the 41st International Conference on
  Machine Learning}}, Vol.~\bibinfo{volume}{235}.
  \bibinfo{pages}{62138--62160}.
\newblock
\urldef\tempurl%
\url{https://proceedings.mlr.press/v235/zhou24r.html}
\showURL{%
\tempurl}


\end{thebibliography}

\end{document}